\documentstyle[12pt]{article}

\advance \topmargin by -\headheight
\advance \topmargin by -\headsep

\evensidemargin \oddsidemargin
\begin{document}

\newpage
\pagestyle{empty}

\begin{flushright}
CERN-TH/99-223\\
gr-qc/9910089
\end{flushright}

\begin{centering}

\baselineskip 21pt plus 0.2pt minus 0.2pt

\bigskip
{\Large {\bf Are~we~at~the~dawn~of 
quantum-gravity~phenomenology?\footnote{Based 
on lectures given at the XXXV Karpacz Winter School of Theoretical Physics
``From Cosmology to Quantum Gravity", Polanica, Poland, 2-12 February, 
1999. To appear in the proceedings.}}}

\bigskip

\baselineskip 12pt plus 0.2pt minus 0.2pt

\bigskip
\bigskip

{\bf Giovanni AMELINO-CAMELIA}\footnote{{\it Marie Curie
Fellow} of the European Union
(address from February 2000: Dipartimento di Fisica,
Universit\'a di Roma ``La Sapienza'',
Piazzale Moro 2, Roma, Italy)}\\
%(permanent address: Dipartimento di Fisica, Universit\'a di&nbsp;
%Roma {\it La Sapienza}, Piazzale Moro 2, Roma, Italy)}\\
\bigskip
Theory Division, CERN, CH-1211, Geneva, Switzerland

\end{centering}
\vspace{1.2cm}

\centerline{\bf ABSTRACT }

\begin{quote}%\baselineskip 13pt

A handful of recent papers has been devoted to 
proposals of experiments capable of testing
some candidate quantum-gravity phenomena.
These lecture notes
emphasize those aspects that are most relevant to
the questions that inevitably
come to mind when one is exposed for the first time
to these research developments:
How come theory and experiments
are finally meeting in spite of all the gloomy forecasts
that pervade traditional quantum-gravity reviews?
Is this a case of theorists
having put forward more and more speculative
ideas until a point was reached
at which conventional experiments could
rule out the proposed phenomena? Or has there been
such a remarkable improvement in experimental
techniques and ideas that we are now capable of testing plausible
candidate quantum-gravity phenomena?
These questions are analysed rather carefully for the 
recent proposals of tests of space-time fuzziness using
modern interferometers and tests of dispersion in the quantum-gravity
vacuum using observations of electromagnetic radiation
from distant astrophysical sources.
I also briefly discuss other 
proposed quantum-gravity experiments, including
those exploiting the properties of the neutral-kaon system 
for tests of quantum-gravity-induced decoherence and those 
using particle-physics accelerators for tests of models with
large extra dimensions.
The emerging picture of ``quantum-gravity phenomenology''
suggests that we are finally starting the exploration
of a relatively large class of
plausible quantum-gravity effects. However, our chances to
obtain positive (discovery) experimental results
still depend crucially on the magnitude of these effects;
in particular,
in most cases the level of sensitivity that the relevant
experiments should achieve within a few years corresponds
to effects suppressed only linearly by the Planck length.

\end{quote}
\baselineskip 18pt

%\medskip\noindent{\bf PACS numbers:}
%11.30.Qc, 12.10.Dm, 14.80.Hv, 90.80.Cq
\vfill

\newpage
\pagenumbering{arabic}
\setcounter{page}{1}
\pagestyle{plain}
\baselineskip 12pt plus 0.2pt minus 0.2pt

\section{INTRODUCTION}

Traditionally the lack of experimental input~\cite{nodata}
has been the most important obstacle 
in the search for ``quantum gravity'', 
the new theory that should provide
a unified description of gravitation and quantum mechanics.
Recently there has been a small, but nonetheless
encouraging, number of  
proposals~\cite{elmn,stringcogwi,peri,gacgrb,ahluexp,
gacgwi,grwlarge,grwlarge2}
of experiments probing the nature of the interplay between 
gravitation and quantum mechanics. At the same time the ``COW-type'' 
experiments on quantum mechanics in a strong (classical)
gravitational environment,
initiated 
by Colella, Overhauser and Werner~\cite{cow},
have reached levels of sensitivity~\cite{chu} such that
even gravitationally induced quantum phases 
due to local tides can be detected.
In light of these developments there 
is now growing (although still understandably
cautious) hope for data-driven 
insight into the structure of quantum gravity.

The primary objective of these lecture notes
is the one of giving the reader an intuitive
idea of how far quantum-gravity phenomenology has come.
This is somewhat tricky. 
Traditionally experimental tests of 
quantum gravity were believed to be not better than a dream.
The fact that now (some) theory and (some) experiments
finally ``meet'' could have two very different explanations:
it could be that experimental
techniques and ideas
have improved so much that now tests of plausible
quantum-gravity effects are within reach,
but it could also be that 
theorists have managed to come
up with scenarios speculative enough
to allow testing by conventional experimental techniques.
I shall argue that experiments have indeed progressed to the point
were some significant quantum-gravity tests are doable.
I shall also clarify in which sense the traditional pessimism
concerning quantum-gravity experiments was built upon the analysis
of a very limited set of experimental ideas, with the
significant omission of the possibility (which we now
find to be within our capabilities) of experiments
set up in such a way that very many of the very small
quantum-gravity effects are somehow summed together.
Some of the theoretical ideas that can be tested experimentally 
are of course quite speculative (decoherence, space-time
fluctuations, large extra dimensions, ...) but this is
not so disappointing because it seems reasonable to expect
that the new theory
should host a large number
of new conceptual/structural
elements in order to be capable of
reconciling the (apparent)
incompatibility between gravitation and quantum mechanics.
[An example of motivation for very new structures
is discussed here in Section~11, which is a ``theory addendum''
reviewing some of the arguments~\cite{gacmpla} 
in support of the idea~\cite{gacgrf98}
that the mechanics on which quantum
gravity is based might not be exactly
the one of ordinary quantum mechanics, since it should
accommodate a somewhat different (non-classical)
concept of ``measuring apparatus''
and a somewhat different
relationship between ``system''
and ``measuring apparatus''.]

In giving the reader an intuitive idea of how
far quantum-gravity phenomenology
has come it will be very useful to rely on simple
phenomenological models of candidate quantum-gravity effects.
The position I am here taking is not that these models should
become cornerstones of theoretical work on quantum gravity
(at best they are possible ways
in which quantum-gravity might manifest itself),
but rather that these models
can be useful in giving an intuitive
description of the
level of sensitivity that experiments are finally reaching.
Depending on the reader's intuition for the quantum-gravity realm
these phenomenological
models might or might not appear likely as faithful
descriptions of effects actually present in quantum-gravity,
but in any case
by the end of these notes the reader should find that these
models are at least
useful for the characterization of the level of sensitivity
that quantum-gravity experiments have reached, and
can also be useful to describe
the progress (past and future) of these sensitivity levels.
In particular, in the ``language'' set up by these models
one can see an emerging picture suggesting that
we are finally ready for the exploration
of a relatively large class of
plausible quantum-gravity effects, even though
our chances to
obtain positive (discovery) experimental results
still depend crucially on the magnitude of these effects:
in most cases the level of sensitivity that the relevant
experiments should achieve within a few years corresponds
to effects suppressed only linearly by the Planck length
$L_p$ ($L_p \sim 10^{-35}m$).

The bulk of these notes gives
brief reviews of the quantum-gravity experiments
that can be done.
The reader will be asked to
forgive the fact that this review is not 
very balanced. The two proposals in which this author has been
involved~\cite{gacgrb,gacgwi}
are in fact discussed in greater detail, 
while for the experiments
proposed in Refs.~\cite{elmn,stringcogwi,peri,grwlarge,grwlarge2}
I just give a very brief discussion 
with emphasis on the most important conceptual ingredients.

The students who attended the School might be surprised to find
the material presented with a completely different strategy.
While my lectures in Polanica were sharply divided 
in a first part on theory and a second part on experiments,
here some of the theoretical intuition is presented while
discussing the experiments. It appears to me that this strategy
might be better suited for a written presentation.
I also thought it might be useful to start with the
conclusions, which are given in the next two sections.
Section 4 reviews the proposal 
of using modern interferometers 
to set bounds on space-time fuzziness.
In Section 5 I review the proposal of using
data on GRBs (gamma-ray bursts)
to investigate possible quantum-gravity 
induced {\it in vacuo} dispersion
of electromagnetic radiation.
In Section 6 I give brief reviews of other 
quantum-gravity experiments.
In Section 7 I give a brief discussion of the
mentioned ``COW-type'' experiments
testing quantum mechanics in a
strong classical-gravity environment.
Section 8 provides a ``theory addendum''
on various scenarios for bounds on the measurability
of distances in quantum gravity and their possible relation
to properties of the space-time foam.
Section 9 provides a theory addendum
on an absolute bound on the measurability
of the amplitude of a gravity wave which should hold even
if distances are not fuzzy.
Section 10 provides a theory addendum
on other works which are in one way or another related to
(or relevant for) the content of these notes.
Section 11 gives the mentioned theory addendum
concerning ideas on a mechanics for quantum gravity
that be not exactly of the type of ordinary quantum mechanics.
Finally in Section 12 I give some comments on the outlook
of quantum-gravity phenomenology, and I also emphasize the fact that,
whether or not they turn out to be helpful for quantum gravity,
most of the experiments considered in these notes are intrinsically
significant as tests of quantum mechanics and/or tests
of fundamental symmetries.

\section{FIRST THE CONCLUSIONS: WHAT HAS THIS PHENOMENOLOGY 
ACHIEVED?}

Let me start with a brief description of the present status
of quantum-gravity phenomenology.
Some of the points made in this section are supported
by analyses which will be reviewed in the following sections.
The crucial question is:
Can we just test some
wildly speculative ideas which have somehow surfaced
in the quantum-gravity literature?
Or can we test even some
plausible candidate quantum-gravity phenomena?

Before answering these questions it is appropriate
to comment on the general expectations we have
for quantum gravity.
It has been realized for some time now
that by combining elements of gravity with
elements of quantum mechanics one is led
to ``interplay phenomena'' with rather distinctive signatures,
such as quantum fluctuations of 
space-time~\cite{wheely,hawk,arsarea},
and violations of Lorentz and/or CPT
symmetries~\cite{ehns,pagegaume,emn,hpcpt,kpoin,thooft,gacxt},
but the relevant effects are expected to be very small (because of the 
smallness of the Planck length).
Therefore in this ``intuition-building'' section the reader must
expect from me a description of experiments with a remarkable 
sensitivity to the new phenomena.

Let me start from the possibility
of quantum fluctuations of space-time.
A prediction of nearly
all approaches to the unification
of gravitation and quantum mechanics is that 
at very short distances the sharp
classical concept of space-time should give way 
to a somewhat ``fuzzy'' (or ``foamy'') picture,
possibly involving virulent
geometry fluctuations (sometimes depicted as
wormholes and black holes popping in and out of the vacuum).
Although the idea of space-time foam remains somewhat vague
and it appears to have significantly
different incarnations in different quantum-gravity approaches,
a plausible expectation that emerges
from this framework is that the distance between two
bodies ``immerged" in the space-time foam
would be affected by (quantum) fluctuations.
If urged to give a rough description of these fluctuations
at present theorists can only guess that they would be 
of Planck-length 
magnitude and occurring at a frequency of roughly one per
Planck time $T_p$ ($T_p = L_p/c \sim 10^{-44} s$).
One should therefore deem significant for space-time-foam
research any experiment that monitors
the distances between two bodies with enough sensitivity
to test this type of fluctuations.
This is exactly what was achieved by the analysis reported
in Refs.~\cite{gacgwi,bignapap}, which was based on the observation
that the most advanced modern interferometers
(the ones normally used for detection of classical gravity waves)
are the natural instruments to study the fuzziness of distances.
While I postpone to Section~4
a detailed discussion of these interferometry-based
tests of fuzziness, let me emphasize 
already here that modern interferometers
have achieved such a level of sensitivity that we are already in a 
position to {\bf rule out} fluctuations in the distances of their 
test masses of the type discussed above, {\it i.e.} fluctuations of
Planck-length magnitude occurring at a rate of one
per each Planck time.
This is perhaps the simplest way for the reader
to picture intuitively
the type of objectives already reached by quantum-gravity
phenomenology.

Another very intuitive measure of the maturity of
quantum-gravity phenomenology comes from the studies
of {\it in vacuo} dispersion proposed in Ref.~\cite{gacgrb}
(also see the more recent purely experimental
analyses \cite{schaef,billetal}).
Deformed dispersion relations are a rather natural
possibility for quantum gravity. For example, they emerge
naturally in quantum gravity scenarios requiring a modification
of Lorentz symmetry. Modifications of 
Lorentz symmetry could result from space-time
discreteness ({\it e.g.} a discrete space
accommodates a somewhat different concept of ``rotation''
with respect to the one of ordinary continuous spaces),
a possibility extensively investigated in the quantum gravity
literature (see, {\it e.g.}, Ref.~\cite{thooft}), 
and it would also naturally result from an ``active'' quantum-gravity
vacuum of the type advocated by Wheeler and Hawking~\cite{wheely,hawk}
(such a ``vacuum''
might physically label the space-time points,
rendering possible the selection
of a ``preferred frame'').
The specific structure of the deformation
can differ significantly from model to model.
Assuming that the deformation admits 
a series expansion at small energies $E$, and parametrizing 
the deformation in terms of an energy\footnote{I 
parametrize deformations of dispersion relations
in terms of an energy scale $E_{QG}$, while I later
parametrize the proposals for distance fuzziness
with a length scale $L_{QG}$.}
scale $E_{QG}$
(a scale characterizing the onset of
quantum-gravity dispersion effects, often identified with
the Planck energy $E_{p} = \hbar c/L_{p} \sim 10^{19} GeV$),
for a massless particle
one would expect to be able to approximate
the deformed dispersion relation at low energies according to
\begin{equation}
c^2{\bf p}^2 \simeq E^2 \left[1 
+ \xi \left({E \over E_{QG}}\right)^{\alpha}
+ O \left( \left( {E \over E_{QG}}\right)^{\alpha+1} \right)
\right]
\label{dispgen}
\end{equation}
where $c$ is the conventional speed-of-light constant.
The scale $E_{QG}$, the power $\alpha$
and the sign ambiguity $\xi = \pm 1$ 
would be fixed in a given dynamical framework;
for example, in some of the approaches
based on dimensionful quantum deformations of
Poincar\'e symmetries~\cite{kpoin,lukipapers,gacmaj}
one encounters a dispersion 
relation $c^2{\bf p}^2 = E_{QG}^2 \, \left[ 1 -
e^{E/E_{QG}}\right]^2$, which implies $\xi = \alpha = 1$.
Because of the smallness of $1/E_{QG}$ it was
traditionally believed that this effect could not be
seriously tested experimentally ({\it i.e.} that,
for $E_{QG} \sim E_p$, experiments would
only be sensitive to values of $\alpha$ much smaller
than $1$), but in Ref.~\cite{gacgrb} it was observed that
recent progress in the phenomenology of GRBs~\cite{grbnew}
and other astrophysical phenomena
should soon allow us
to probe values of $E_{QG}$ of the order
of (or even greater than) $E_p$ for values of $\alpha$
as large as $1$. As discussed later in these
notes, $\alpha = 1$ appears to be the smallest value
that can be obtained with plausible quantum-gravity arguments
and several of these arguments actually
point us toward the larger value $\alpha = 2$,
which is still very far from present-day experimental
capabilities. While of course it would be very important
to achieve sensitivity to
both the $\alpha = 1$ and the $\alpha = 2$ scenarios,
the fact that we will soon test $\alpha = 1$
is a significant first step.

Another recently proposed quantum-gravity experiment
concerns possible violations of CPT invariance.
This is a rather general prediction of quantum-gravity
approaches, which for example can be due to elements of
nonlocality 
(locality is one of the hypotheses of the ``CPT theorem'')
and/or elements of decoherence
present in the approach.
At least some level of non-locality is quite natural for
quantum gravity as a theory with a natural length scale
which might play the role of ``minimum
length''~\cite{padma,venekonmen,dharam94grf,gacmpla,garay}.
Motivated by the structure of ``Liouville
strings''~\cite{emn} (a non-critical string approach to
quantum gravity which appears to admit a space-time foam
picture)
a phenomenological parametrization of
quantum-gravity induced CPT violation in
the neutral-kaon system
has been proposed in Refs.~\cite{ehns,emnk}.
(Other studies of the phenomenology of CPT violation
can be found in Ref.~\cite{hpcpt,floreacpt}.)
In estimating the parameters
that appear in this phenomenological
model the crucial point is as usual the overall
suppression given by some power of the Planck
length $L_p \sim 1/E_p$.
For the case in which the Planck length enters only
linearly in the relevant formulas,
experiments investigating the properties
of neutral kaons
are already setting significant bounds
on the parameters of this phenomenological
approach~\cite{elmn}.

In summary, experiments are reaching significant sensitivity
with respect to all of the frequently discussed features
of quantum gravity that I mentioned at the beginning of this
section: space-time fuzziness, violations of Lorentz invariance,
and violations of CPT invariance.
Other quantum-gravity experiments, which I shall discuss
later in these notes, can probe other
candidate quantum-gravity phenomena, giving additional breadth
to quantum-gravity phenomenology.

Before closing this section there is one more answer I should
give: how could this happen in spite of all the gloomy
forecasts
which one finds in most quantum-gravity review papers?
The answer is actually simple. Those gloomy forecasts were based
on the observation that under ordinary conditions
the direct detection of a single quantum-gravity phenomenon
would be well beyond our capabilities
if the magnitude of the phenomenon is suppressed by
the smallness of the Planck length.
For example, in particle-physics contexts
this is seen in the fact that
the contribution from ``gravitons''
(the conjectured mediators of quantum-gravity interactions)
to particle-physics
processes with center-of-mass energy ${\cal E}$
is expected to
be penalized by
overall factors given by some power of the
ratio ${\cal E}/(10^{19} GeV)$.
However, small effects can become observable in
special contexts and in particular one can always
search for an experimental setup such that
a very large number of the very small quantum-gravity
contributions are effectively summed together.
This later possibility is not unknown to the particle-physics
community, since it has been
exploited in the context of investigations
of the particle-physics theories unifying
the strong and electroweak interactions, were one encounters
the phenomenon of proton decay.
By keeping under observation very large numbers
of protons, experimentalists have
managed\footnote{This author's familiarity~\cite{gactesi}
with the accomplishments of proton-decay experiments has
certainly contributed to the moderate optimism for the outlook
of quantum-gravity phenomenology which is found in these
notes.}
to set highly significant
bounds on proton decay~\cite{proton},
even though the proton-decay probability
is penalized by the fourth power of the small ratio between
the proton mass, which is of order $1 GeV$, and 
the mass of the vector bosons expected to
mediate proton decay,
which is conjectured to be of order $10^{16} GeV$.
Just like proton-decay experiments are based on a simple
way to put together very many of the small proton-decay
effects\footnote{For each of the protons being monitored 
the probability of decay is extremely small, but
there is a significantly large probability that 
at least one of the many monitored protons decay.}
the experiments using modern interferometers
to study space-time fuzziness and the experiments
using GRBs to study violations of Lorentz invariance
exploit simple ways to put together very many of the
very small quantum-gravity effects. I shall explain this
in detail in Sections 4 and 5.

\section{ADDENDUM TO CONCLUSIONS: ANY HINTS TO THEORISTS
FROM EXPERIMENTS? }

In the preceding section I have argued that quantum-gravity
phenomenology, even being as it is in its infancy, is already
starting to provide the first significant tests of plausible
candidate quantum-gravity phenomena.
It is of course just ``scratching the surface''  of
whatever ``volume'' contains the full collection of experimental
studies we might wish to perform, but we are finally
getting started.
Of course, a phenomenology programme is meant to provide
input to the theorists working in the area,
and therefore one measure
of the achievements of a phenomenology programme is given by
the impact it is having on theory studies.
In the case of quantum-gravity experiments the flow of information
from experiments to theory will take some time. The primary reason
is that most quantum-gravity approaches have been guided
(just because there was no alternative guidance from data)
by various sorts of
formal intuition for quantum gravity
(which of course remain pure speculations
as long as they are not confirmed by experiments).
This is in particular true for the two
most popular approaches to the unification of gravitation
and quantum
mechanics, {\it i.e.} ``critical
superstrings''~\cite{dbrane,critstring}
and ``canonical/loop quantum gravity''~\cite{canoloop}.
Because of the type of intuition that went into them,
it is not surprising that these ``formalism-driven
quantum gravity approaches'' are proving extremely useful
in providing us new ideas on how gravitation
and quantum mechanics could resolve the apparent conflicts
between their conceptual structures,
but they are not giving us any ideas on which experiments
could give insight into the nature of quantum gravity.
The hope that these approaches could
eventually lead to new intuitions for the nature of space-time
at very short distances has been realized
only rather limitedly.
In particular, it is still unclear if and how
these formalisms host the mentioned scenarios
for quantum fluctuations of 
space-time and violations of Lorentz
and/or CPT symmetries.
The nature of the quantum-gravity vacuum (in the sense discussed
in the preceding section)
appears to be still very far ahead in the
critical superstring research programme and its analysis is only
at a very preliminary stage within canonical/loop quantum gravity.
In order for the experiments discussed in these notes to
affect directly critical superstring research
and research in canonical/loop quantum gravity
it is necessary to make substantial progress in the analysis
of the physical implications of these formalisms.

Still, in an indirect way the recent results of quantum-gravity
phenomenology have already started to have an impact
on theory work in these
quantum gravity approaches. The fact that it is becoming clear
that (at least a few) quantum-gravity experiments can be done
has reenergized efforts to explore the
physical implications of the formalisms. The best example of
this way in which phenomenology can influence ``pure theory''
work is provided by Ref.~\cite{gampul},
which was motivated by the results reported in Ref.~\cite{gacgrb}
and showed that 
canonical/loop quantum gravity
admits (under certain conditions, which in particular
involve some parity breaking)
the phenomenon of deformed dispersion
relations, with deformation going linearly with
the Planck length.

While the impact on theory work in the formalism-driven
quantum gravity approaches is still quite limited,
of course the new experiments are providing
useful input
for more intuitive/phenomelogical theoretical
work on quantum gravity.
For example, the analysis reported in Refs.~\cite{gacgwi,bignapap},
by ruling out the scheme of distance fluctuations of
Planck length magnitude occurring at a rate of one per
Planck time, has had significant impact~\cite{bignapap,nggwi}
on the line of research which has been deriving
intuitive pictures of properties of quantum space-time
from analyses of measurability and uncertainty
relations~\cite{gacmpla,karo,diosi,ng}.
Similarly the ``Liouville-string''~\cite{emn}
inspired phenomenological approach to quantum
gravity~\cite{emnk,aemn1}
has already received important input from
the mentioned studies of the neutral-kaon system
and will receive equally important input from
the mentioned GRB experiments,
once these experiments (in a few years)
reach Planck-scale sensitivity.

\section{INTERFEROMETRY AND FUZZY SPACE-TIME}

In the preceding two sections I have described the conclusions 
which I believe
to be supported by the present status of quantum-gravity
phenomenology.
Let me now start providing some support for those conclusions
by reviewing my proposal \cite{gacgwi,bignapap}
of using modern interferometers to set bounds on space-time fuzziness.
I shall articulate this in subsections because some preliminaries
are in order. Before going to the analysis of experimental data
it is in fact necessary to give
a proper (operative) definition of fuzzy distance
and give a description of the type of stochastic properties 
one might expect of quantum-gravity-induced
fluctuations of distances.

\subsection{Operative definition of fuzzy distance}

While nearly all approaches to the unification
of gravity and quantum mechanics appear to lead
to a somewhat fuzzy
picture of space-time, 
within the various formalisms it is often difficult to
characterize physically this fuzziness.
Rather than starting from formalism, I shall advocate
an operative definition of fuzzy space-time.
More precisely for the time being I shall just consider 
the concept of fuzzy distance.
I shall be guided by the expectation that
at very short distances the sharp
classical concept of distance should give way 
to a somewhat fuzzy distance. Since interferometers
are ideally suited to monitor the distance between
test masses, I choose as operative definition of 
quantum-gravity-induced fuzziness 
one which is expressed in terms
of quantum-gravity-induced noise in the 
read-out of interferometers.

In order to properly discuss this proposed definition
it will prove useful
to briefly review some aspects of the physics of 
modern Michelson-type interferometers.
These are schematically composed~\cite{saulson}
of a (laser) light source, a beam splitter 
and two fully-reflecting mirrors placed
at a distance
$L$ from the beam splitter in orthogonal directions.
The light beam is decomposed by the beam splitter 
into a transmitted beam directed toward one of the mirrors
and a reflected beam directed toward the other mirror;
the beams are then reflected by the mirrors 
back toward the beam splitter,
where~\cite{saulson} they are 
superposed\footnote{Although all modern interferometers
rely on the technique of folded interferometer's arms
(the light beam bounces several times between the
beam splitter and the mirrors before superposition),
I shall just discuss 
the simpler ``no-folding'' conceptual setup.
The readers familiar with the subject can easily realize
that the observations here reported also apply
to more realistic setups, although in some steps of the derivations
the length $L$ would have to be understood as the optical length
(given by the actual length of the arms multiplied by the number
of foldings).}.
The resulting interference pattern 
is extremely sensitive to changes in the positions of the mirrors
relative to the beam splitter.
The achievable
sensitivity is so high that planned interferometers~\cite{ligo,virgo}
with arm lengths $L$ of $3$ or $4$ $Km$
expect to detect gravity waves of amplitude $h$ as 
low as $3 \cdot 10^{-22}$ at frequencies of about $100 Hz$.
This roughly means that these modern gravity-wave interferometers
should monitor the (relative) positions of their test masses
(the beam splitter and the mirrors)
with an accuracy~\cite{ligoprototype}
of order $10^{-18} m$ and better.

In achieving this remarkable accuracy experimentalists must
deal with classical-physics displacement noise sources ({\it e.g.},
thermal and seismic effects induce fluctuations in the relative
positions of the test masses) and displacement noise sources
associated to effects of ordinary quantum mechanics
({\it e.g.}, the combined minimization
of {\it photon shot noise} and {\it radiation pressure noise}
leads to an irreducible noise source which has its root in
ordinary quantum mechanics~\cite{saulson}).
The operative definition of fuzzy distance which I advocate
characterizes the corresponding quantum-gravity effects
as an additional source of displacement noise.
A theory in which the concept of distance is 
fundamentally fuzzy in this 
operative sense would be such that even in the idealized
limit in which all classical-physics and ordinary-quantum-mechanics
noise sources are completely eliminated the read-out of an
interferometer would still be noisy as a result of
quantum-gravity effects.

Upon adopting this operative definition of fuzzy distance,
interferometers are of course the natural tools for 
experimental tests of proposed distance-fuzziness scenarios.

I am only properly discussing distance fuzziness
although ideas on space-time foam
would also motivate investigations of time fuzziness.
It is not hard to modify the definition here advocated
for distance fuzziness to describe time fuzziness
by replacing the interferometer
with some device that keeps track of the
synchronization of a pair of clocks\footnote{Actually,
a realistic analysis of ordinary Michelson-type 
interferometers is likely to lead to a contribution
from space-time foam to noise levels that is the sum
(in some appropriate sense) of the effects due
to distance fuzziness and time fuzziness ({\it e.g.} associated
to the frequency/time measurements involved).}.
I shall not pursue this matter further since
I seem to understand\footnote{This
understanding is mostly based on recent conversations
with G.~Busca and P.~Thomann who are involved
in the next generation of ultra-precise clocks
to be realized in microgravity (outer space) environments.}
that sensitivity to time fluctuations is still significantly
behind the type of sensitivity to distance fluctuations
achievable with modern Michelson-type experiments.

\subsection{Random-walk noise from random-walk models
of quantum space-time fluctuations}

As already mentioned in Section 2,
it is plausible 
that a quantum space-time
might involve fluctuations of magnitude $L_{p}$
occurring at a rate of roughly one per each time
interval of magnitude $t_{p} = L_{p}/c \sim 10^{-44} s$.
One can start investigating this scenario by
considering the possibility that experiments monitoring
the distance $D$ between two bodies for a time $T_{obs}$
(in the sense appropriate, {\it e.g.}, for an interferometer)
could involve a total effect amounting
to $n_{obs} \equiv T_{obs}/t_{p}$ 
randomly directed fluctuations\footnote{One might actually
expect even more than $T_{obs}/t_{p}$ fluctuations of
magnitude $L_p$ in a time $T_{obs}$
depending on how frequent fluctuations occur in the
region of space spanned by the distance $D$.
This and other possibilities will be later modelled
by replacing $L_p$ with a phenomenological scale $L_{QG}$
which could even depend on $D$.
However, as mentioned in the Introduction,
rather than focusing on
the details of the physics
of the fuzziness models, I am here
discussing models from the point of view of a
characterization of the levels of quantum-gravity sensitivity
reached by recent experiments,
and the scale $L_{QG}$ will be seen primarily from this perspective
rather than attempting careful estimates in terms of one or
another picture of space-time fluctuations.}
of magnitude $L_{p}$.
An elementary analysis
allows to establish that in such a context
the root-mean-square deviation $\sigma_D$ would be
proportional to $\sqrt{T_{obs}}$:
\begin{equation}
\sigma_D \sim  \sqrt{c L_{p} T_{obs}} \, .
\label{no1bis}
\end{equation}

{}From the type of $T_{obs}$-dependence of Eq.~(\ref{no1bis})
it follows~\cite{gacgwi} that the corresponding quantum fluctuations 
should have displacement amplitude spectral density $S(f)$
with the $f^{-1}$
dependence\footnote{Of course,
in light of the nature of the arguments used,
one expects that
an $f^{-1}$ dependence of the quantum-gravity induced $S(f)$
could only be valid for frequencies $f$ significantly
smaller than the Planck frequency $c/L_{p}$
and significantly larger than the inverse of the time scale
over which, even ignoring the gravitational field
generated by the devices, the classical geometry 
of the space-time region where
the experiment is performed manifests
significant curvature effects.}
typical of ``random walk noise''~\cite{rwold}:
\begin{eqnarray}
S(f) = f^{-1} \sqrt {c \, L_{p}} ~.
\label{gacspectrforlp}
\end{eqnarray}
In fact, there is a general connection
between $\sigma  \sim \sqrt{T_{obs}}$
and $S(f) \sim f^{-1}$,
which follows~\cite{rwold} from the
general relation
\begin{eqnarray}
\sigma^2 = \int_{1/T_{obs}}^{f_{max}}
[S(f)]^2 \,  df ~,
\label{gacspectrule}
\end{eqnarray}
valid for a frequency band limited from below only
by the time of observation $T_{obs}$.

The displacement amplitude spectral density 
(\ref{gacspectrforlp}) provides a very useful characterization
of the random-walk model
of quantum space-time fluctuations prescribing
fluctuations of magnitude $L_{p}$
occurring at a rate of roughly one per 
each time interval of magnitude $L_{p}/c$.
If somehow we have been assuming the wrong magnitude of 
distance fluctuations
or the wrong rate (also see Subsection 4.4)
but we have been correct in taking a random-walk model
of quantum space-time fluctuations Eq.~(\ref{gacspectrforlp})
should be replaced by
\begin{eqnarray}
S(f) = f^{-1} \sqrt {c \, L_{QG}} ~,
\label{gacspectr}
\end{eqnarray}
where $L_{QG}$ is the appropriate length scale
that takes into account the correct values of magnitude and rate of 
the fluctuations.

If one wants to be open to the possibility
that the nature of the stochastic processes
associated to quantum space-time be not exactly 
(also see Section 8)
the one of a random-walk model
of quantum space-time fluctuations, then the $f$-dependence 
of the displacement amplitude spectral density 
could be different. This leads one to consider 
the more general parametrization
\begin{eqnarray}
S(f)=f^{-\beta} \, c^{\beta-{1 \over 2}} \, 
({\cal L}_{\beta})^{{3 \over 2}-\beta} ~.
\label{gacspectrbeta}
\end{eqnarray}
In this general parametrization the
dimensionless quantity $\beta$
carries the information on the nature of the underlying stochastic
processes, while the length scale ${\cal L}_{\beta}$ carries
the information on the magnitude and rate of the fluctuations.
I am assigning an
index $\beta$ to ${\cal L}_{\beta}$ just in order to facilitate
a concise description of experimental bounds; for example,
if the fluctuations scenario with, say, $\beta = 0.6$
was ruled out down to values of the effective length
scale of order, say, $10^{-27}m$ I would just
write ${\cal L}_{\beta= 0.6} < 10^{-27}m$. 
As I will discuss in Section 8, one might be interested in 
probing experimentally all values of $\beta$ in the 
range $1/2 \le \beta \le 1$, with special interest in the 
cases $\beta = 1$ (the case of random-walk models
whose effective length scale I denominated
with $L_{QG} \equiv {\cal L}_{\beta=1}$), $\beta = 5/6$,
and $\beta = 1/2$.

\subsection{Comparison with gravity-wave interferometer data}

Before discussing experimental bounds
on ${\cal L}_{\beta}$ from gravity-wave interferometers,
let us fully appreciate the significance of these bounds
by getting some intuition on the actual magnitude of the
quantum fluctuations I am discussing.
One intuition-building observation is that even
for the case $\beta = 1$, which among the cases I consider is 
the one with the most virulent
space-time fluctuations, the fluctuations predicted are truly
minute: the $\beta = 1$ relation (\ref{no1bis})
only predicts fluctuations with standard deviation of
order $10^{-5}m$ on a time of observation as large as
$10^{10}$ years (the size of the whole observable universe
is about $10^{10}$ light years!!).
In spite of the smallness of these effects,
the precision~\cite{saulson}
of modern interferometers
(the ones whose primary objective is the detection
of the classical-gravity phenomenon of gravity waves)
is such that we can obtain significant information at least
on the scenarios with values of $\beta$ toward the high
end\footnote{As mentioned, for $L_{QG} = L_p$
the case $\beta = 1$ corresponds to
a mean-square deviation induced by
the distance fluctuations that is only linearly suppressed
by $L_p$: $\sigma^2_D \sim L_p c T$.
Analogously, values of $\beta$ in the
interval $1/2 < \beta < 1$ correspond to
$\sigma^2_D$ suppressed by a power of $L_p$ between 1 and 2.
The fact that we can only test values of $\beta$ 
toward the high end
of the interval $1/2 \le \beta \le 1$
can be intuitively characterized by stating that
the fuzziness models we are able to test
have $\sigma^2_D$ that is not much
more than linearly suppressed by the Planck length.}
of the interesting interval $1/2 \le \beta \le 1$,
and in particular we can investigate quite sensitively
the intuitive case of the random-walk model of
space-time fluctuations.
The operation of gravity-wave interferometers 
is based on the detection of
minute changes in the positions of some test masses
(relative to the position of a beam splitter).
If these positions were affected by
quantum fluctuations of the type discussed
above, the operation of gravity-wave interferometers
would effectively involve an additional
source of noise due to quantum gravity.

This observation allows to set interesting bounds already
using existing noise-level data obtained at
the {\it Caltech 40-meter interferometer}, which
has achieved~\cite{ligoprototype}
displacement noise levels with amplitude spectral density
lower than $10^{-18} m/\sqrt{H\!z}$
for frequencies between $200$ and $2000$ $H\!z$.
While this is still very far from the levels
required in order to probe significantly the lowest values
of $\beta$ (for ${\cal L}_{\beta = 1/2} \sim L_{p}$
and $f \sim 1000 Hz$ the quantum-gravity noise induced in the
$\beta = 1/2$ scenario
is only of order $10^{-36} m/\sqrt{H\!z}$), 
these sensitivity levels clearly rule out all values
of $L_{QG}$ ({\it i.e.} ${\cal L}_{\beta = 1}$)
down to the Planck length.
Actually, even values of $L_{QG}$ significantly
smaller than the Planck length are inconsistent with the data 
reported in Ref.~\cite{ligoprototype}; in particular,
from the observed noise level of $3 \cdot 10^{-19} m/\sqrt{H\!z}$
near $450$ $H\!z$, which is the best achieved
at the {\it Caltech 40-meter interferometer},
one obtains~\cite{gacgwi}
the bound $L_{QG} \le 10^{-40}m$.
As discussed above,
the random-walk
model of distance fuzziness
with fluctuations of magnitude $L_{p}$
occurring at a rate of one per each $t_{p}$ time interval,
would correspond to the
prediction $L_{QG} \sim  L_p \sim 10^{-35}m$
and it is therefore {\bf ruled out by these data}.
This experimental information implies that,
if one was to insist on this type of models,
realistic random-walk models 
of quantum space-time fluctuations 
would have to be significantly
less noisy (smaller prediction for $L_{QG}$)
than the intuitive 
one which is now ruled out.
Since, as I shall discuss, there are rather plausible
scenarios for significantly less noisy random-walk models,
it is important that experimentalists keep pushing forward
the bound on $L_{QG}$.
More stringent bounds on $L_{QG}$
are within reach of 
the LIGO/VIRGO~\cite{ligo,virgo}
generation of gravity-wave interferometers.\footnote{Besides
allowing an improvement on the bound on $L_{QG}$ intended
as a universal property of quantum gravity,
the LIGO/VIRGO generation of interferometers
will also allow us to explore the idea that $L_{QG}$
might be a scale that depends on the experimental context
in such a way that larger interferometers pick up more
of the space-time fluctuations. Based on the intuition
coming from the Salecker-Wigner limit (here reviewed in Section~8),
or just simply on
phenomenological models in which distance fluctuations affect
equally each $L_p$-long segment of a given distance,
it would not be surprising if $L_{QG}$ was a growing function
of the length of the arms of the interferometer.
This gives added significance to the step from the 40-meter
arms of the existing Caltech interferometer
to the few-Km arms of LIGO/VIRGO interferometers.}

In planning future experiments, possibly taylored to test
these effects (unlike LIGO and VIRGO which
were tailored around the properties needed for gravity-wave
detection),
it is important to observe that
an experiment achieving 
displacement noise levels with amplitude spectral density $S^*$
at frequency $f^*$ will set a bound on ${\cal L}_\beta$ of order
\begin{eqnarray}
{\cal L}_{\beta} < \left[ S^* \, (f^*)^\beta \,
c^{(1-2\beta)/2} \right]^{2/(3 - 2 \beta)} ~,
\label{futureboundbeta}
\end{eqnarray}
which in particular for random-walk models takes the form
\begin{eqnarray}
{\cal L}_{\beta} < \left[ {S^* \, f^*  \over
\sqrt{c}} \right]^{2} ~.
\label{futureboundbetarw}
\end{eqnarray}
The structure of Eq.~(\ref{futureboundbeta})
(and Eq.~(\ref{futureboundbetarw})) shows that there can be instances
in which a very large interferometer (the ideal tool for
low-frequency studies) might not be better than a smaller
interferometer, if the smaller one achieves a very small
value of $S^*$.

The formula (\ref{futureboundbeta}) can also be used to
describe as a function of $\beta$
the bounds on ${\cal L}_{\beta}$ achieved by the data
collected at the {\it Caltech 40-meter interferometer}.
Using again the
fact that a noise level of
only $S^* \sim 3 \cdot 10^{-19} m/\sqrt{H\!z}$
near $f^* \sim 450$ $H\!z$ was achieved~\cite{ligoprototype},
one obtains the bounds
\begin{eqnarray}
[{\cal L}_{\beta}]_{caltech}
< \left[ {3 \cdot 10^{-19} m \over \sqrt{H\!z}}
 \, (450 H\!z)^\beta \,
c^{(1-2\beta)/2} \right]^{2/(3 - 2 \beta)} ~.
\label{boundcalty}
\end{eqnarray}

Let me comment in particular on the case $\beta = 5/6$
which might deserve special attention because of its
connection (which was derived in Refs.~\cite{gacgwi,bignapap}
and will be reviewed here in Section~8)
with certain arguments for bounds on the measurability
of distances in quantum gravity~\cite{bignapap,ng,karo}.
From Eq.~(\ref{boundcalty}) we find that ${\cal L}_{\beta=5/6}$
is presently bound to the
level ${\cal L}_{\beta=5/6} \le 10^{-29} m$.
This bound is remarkably stringent in absolute terms, but is still
quite far from the range of values one ordinarily considers
as likely candidates for length scales appearing in quantum gravity.
A more significant bound on ${\cal L}_{\beta=5/6}$
should be obtained by the LIGO/VIRGO generation 
of gravity-wave interferometers.
For example, it is plausible~\cite{ligo} that 
the ``advanced phase'' of LIGO achieve a displacement noise spectrum 
of less than $10^{-20} m/ \sqrt{H\!z}$ near $100$ $H\!z$
and this would probe values of ${\cal L}_{\beta=5/6}$ as small
as $10^{-34} m$.

Interferometers are our best long-term hope for the 
development of this phenomenology, and that 
is why the analysis
in this Section focuses on interferometers. However,
it should be noted that among detectors already in operation
the best bound on $L_\beta$ (if taken as universal)
comes from resonant-bar detectors 
such as NAUTILUS~\cite{nautilus}, which achieved
displacement sensitivity of about $10^{-21}m/\sqrt{H\!z}$ 
near $924 H\! z$.
Correspondingly, one obtains the bound
\begin{eqnarray}
[{\cal L}_{\beta}]_{bars}
< \left[ {10^{-21} m \over \sqrt{H\!z}}
 \, (924 H\!z)^\beta \,
c^{(1-2\beta)/2} \right]^{2/(3 - 2 \beta)} ~.
\label{boundnauti}
\end{eqnarray}

In closing this subsection on interferometry data analysis relevant
for space-time fuzziness scenarios, let me clarify how it happened
that such small effects could be tested.
As I already mentioned, one of the viable strategies
for quantum-gravity experiments is the one finding ways to put
together very many of the very small quantum-gravity effects.
In these interferometric studies
that I proposed in Ref.~\cite{gacgwi}
one does indeed effectively sum up a large number of
quantum space-time fluctuations.
In a time of observation as long as the inverse of the 
typical gravity-wave interferometer frequency of operation
an extremely large number of minute quantum fluctuations
could affect the distance between the test masses.
Although these fluctuations average out, they do leave 
traces in the interferometer. These traces grow with the time
of observation: the standard deviation increases
with the time of observation, while the
displacement noise amplitude spectral density increases
with the inverse of frequency (which again
effectively means that it increases
with the time of observation).
From this point of view it is not surprising that plausible
quantum-gravity scenarios ($1/2 \le \beta \le 1$)
all involve higher noise at lower frequencies: the observation
of lower frequencies requires longer times and is therefore
affected by a larger number of quantum-gravity fluctuations.

\subsection{Less noisy random-walk models
of distance fluctuations?}

The most intuitive result obtained
in Refs.~\cite{gacgwi,bignapap}
and reviewed in the preceding subsection is that
we can rule out the picture in which the
distances between the test masses of the interferometer
are affected by fluctuations of magnitude $L_{p}$
occurring at a rate of one per each $t_{p}$ time interval.
Does this rule out completely the possibility
of a random-walk model of distance fluctuations?
or are we just learning that the most intuitive/naive
example of such a model does not work, but there are other
plausible random-walk models?

Without wanting to embark on a discussion of the
plausibility of less noisy random-walk models,
I shall nonetheless discuss some ideas which could
lead to this noise reduction.
Let me start by observing that certain studies
of measurability of distances in quantum gravity
(see Ref.~\cite{bignapap} and the brief review of those
arguments which is provided in parts of Section~8)
can be interpreted as suggesting that $L_{QG}$
might not be a universal length scale,
{\it i.e.} it might depend on some specific properties of the 
experimental setup 
(particularly the energies/masses involved),
and in some cases $L_{QG}$ could
be significantly smaller than $L_{p}$.

Another possibility one might want to consider~\cite{bignapap}
is the one in
which the quantum properties of space-time are such that 
fluctuations of magnitude $L_{p}$ would occur with frequency
somewhat lower than $1/t_{p}$.
This might happen for various reasons,
but a particularly intriguing possibility\footnote{This possibility
emerged in discussions with Gabriele Veneziano.
In response to my comments on the possibility of fluctuations with 
frequency somewhat lower than $1/t_{p}$ Gabriele
made the suggestion that extended fundamental objects might 
be less susceptible than point particles
to very localized space-time fluctuations.
It would be interesting to work out
in some detail an example of dynamical model of strings in
a fuzzy space-time.}
is the one of theories whose fundamental objects are not
pointlike, such as the popular string theories.
For such theories it is plausible that fluctuations occurring
at the Planck-distance level might have only a modest impact
on extended fundamental objects characterized by a length
scale significantly larger than the Planck length ({\it e.g.}
in string theory the string size, or ``length'', might be
an order of magnitude larger than the Planck length).
This possibility is interesting in general
for quantum-gravity
theories with a hierarchy of length scales, such as
certain ``M-theory motivated'' scenarios with an extra length
scale associated to the compactification from 11 to 10 dimensions.

Yet another possibility for a random-walk model to cause
less noise in interferometers could emerge if somehow
the results of the schematic analysis adopted here and
in Refs.~\cite{gacgwi,bignapap} turned out to be significantly
modified once we become capable of handling all of the details
of a real interferometer.
To clarify which
type of details
I have in mind 
let me mention
as an example the fact that in my analysis the structure
of the test masses was not taken into account in any way:
they were essentially treated as point-like.
It would not be too surprising if
we eventually became able to construct theoretical models
taking into account the interplay between
the binding forces that keep together (``in one piece'')
a macroscopic test mass as well as some random-walk-type
fundamental fluctuations of the space-time in which
these macroscopic bodies ``live''.
The interference pattern observed in the laboratory
reflects the space-time fluctuations only filtered
through their interplay with the mentioned binding forces
of the macroscopic test masses.
These open issues are certainly important and
a lot of insight could be gained through
their investigation,
but there is also some confusion that might
easily result\footnote{In particular,
as I emphasized in Ref.~\cite{swok}, 
these and other elements of confusion
are responsible for the incorrect conclusions
on the Salecker-Wigner measurability limit which
were drawn in
the very recent Ref.~\cite{anos}.}
from simple-minded considerations
(possibly guided by intuition
developed using rudimentary table-top interferometers)
concerning the macroscopic nature of the test masses
used in modern interferometers. 
In closing this section let me try to offer
a few relevant clarifications.
I need to start by adding some comments
on the stochastic processes I have been considering.
In most physical contexts a series of random steps does not
lead to $\sqrt{T_{obs}}$ dependence of $\sigma$
because often the context is such that through the
fluctuation-dissipation theorem
the source of $\sqrt{T_{obs}}$ dependence is
(partly) compensated (some sort of restoring effect).
The hypothesis explored in these discussions of random-walk
models of space-time fuzziness
is that the type of underlying dynamics of quantum space-time
be such that the fluctuation-dissipation theorem be satisfied
without spoiling the $\sqrt{T_{obs}}$ dependence of $\sigma$.
This is an intuition which apparently is shared by other
authors; for example, the study reported in
Ref.~\cite{garayclock}
(which followed by a few months Ref.~\cite{gacgwi},
but clearly was the result of completely independent work)
also models some implications of quantum space-time
(the ones that affect clocks) with stochastic processes
whose underlying dynamics does not produce any dissipation
and therefore the ``fluctuation contribution'' to
the $T_{obs}$ dependence is left unmodified,
although the fluctuation-dissipation theorem
is fully taken into account.

Since a mirror of an interferometer of LIGO/VIRGO type 
is in practice an extremity
of a pendulum, another aspect that the reader might at first
find counter-intuitive is
the fact that the $\sqrt{T_{obs}}$ dependence
of $\sigma$, although coming in with a very small prefactor,
for extremely large $T_{obs}$ would seem to give values of $\sigma$
too large to be consistent
with the structure of a pendulum.
This is a misleading intuition which originates from the experience
with ordinary (non-quantum-gravity) analyses of the pendulum.
In fact, the dynamics of an ordinary pendulum
has one extremity ``fixed'' to a very heavy macroscopic
and rigid body,
while the other extremity is fixed to a much lighter
(but, of course, still macroscopic) body.
The usual stochastic processes considered in
the study of the pendulum affect the heavier body
in a totally negligible way, while they have strong impact
on the dynamics of the lighter body.
A pendulum analyzed according to a random-walk model
of space-time fluctuations
would be affected by stochastic processes which are of the
same magnitude both for its heavier and its lighter extremity.
[The bodies are fluctuating along with the intrinsic
space-time fluctuations, rather than fluctuating as a result of,
say, collisions with material particles occurring in a conventional
space-time.]
In particular, in the directions orthogonal to
the vertical axis the stochastic processes affect the
position of the center of mass of the entire pendulum
just as they would affect the position of the center of
mass of any other body (the spring that connects the two
extremities of the pendulum would not affect the motion of
the overall center of mass of the pendulum).
With respect to the application of some of these
considerations to modern gravity-wave interferometers
it is also important to keep in mind that the measurement strategy
of these interferometers requires that their test masses be 
free-falling.

\section{GAMMA-RAY \space
BURSTS \space AND \space IN-VACUO \space $~$  \space $~$
DISPERSION}

Let me now discuss the proposal put forward in
Ref.~\cite{gacgrb} (also see Ref.~\cite{aemns2}),
which exploits the recent confirmation that at least some
gamma-ray bursters are indeed at cosmological
distances~\cite{paradis,groot,metzeger1,metzeger2},
making it possible for
observations of these to provide interesting
constraints on the fundamental laws of physics.
In particular, such cosmological distances
combine with the short time structure seen in emissions
from some GRBs~\cite{review} to provide ideal features for tests 
of possible {\it in vacuo} dispersion
of electromagnetic radiation from GRBs,
of the type one might expect based on the intuitive
quantum-gravity arguments reviewed in Section~2.
As mentioned, a quantum-gravity-induced
deformation of the dispersion relation for photons
would naturally take the 
form $c^2{\bf p}^2 = E^2 \,[1+ {\cal F} (E/E_{QG})]$,
where $E_{QG}$ is an effective quantum-gravity
energy scale and ${\cal F}$ is a model-dependent function of the
dimensionless ratio $E/E_{QG}$.  In quantum-gravity scenarios in
which the Hamiltonian equation
of motion ${\dot x}_i\,=\partial\,H/\partial\,p_i$ is still valid
(at least approximately valid; valid to an extent sufficient to justify
the analysis that follows)
such a deformed
dispersion relation would lead to energy-dependent velocities
for massless particles, with implications for the
electromagnetic signals that we receive from astrophysical
objects at large distances.
At small energies $E \ll\,E_{QG}$,
it is reasonable to expect that
a series expansion of the dispersion relation should be applicable
leading to the formula (\ref{dispgen}).
For the case $\alpha = 1$, which is the most optimistic
(largest quantum-gravity effect) among the cases discussed
in the quantum-gravity literature,
the formula (\ref{dispgen}) reduces to
\begin{equation}
c^2{\bf p}^2 \simeq E^2 \left(1 
+ \xi {E \over E_{QG}}
\right) ~.
\label{dispauno}
\end{equation}
Correspondingly one would predict
the energy-dependent velocity formula
\begin{eqnarray}
 v = {\partial E \over \partial p} \sim
 c  \left( 1 - \xi {E \over E_{QG}} \right) ~.
\label{vdef}
\end{eqnarray}
To elaborate a bit more than I did in Section~2 on the intuition
that leads to this type of candidate quantum-gravity effect
let me observe that~\cite{gacgrb}
velocity dispersion such as described in (\ref{vdef})
could result from a picture of the vacuum
as a quantum-gravitational `medium', which responds differently to
the propagation of particles of different energies and hence
velocities. This is analogous to propagation through a conventional
medium, such as an electromagnetic plasma~\cite{latorre}.  The
gravitational `medium' is generally believed to contain microscopic
quantum fluctuations,
such as the ones considered in the previous sections.
These may~\cite{garaythermal} be
somewhat analogous to the thermal fluctuations in a
plasma, that occur on time scales of order $t \sim 1 / T$, where $T$
is the temperature.  Since it is a much `harder' phenomenon associated
with new physics at an energy scale far beyond typical photon
energies, any analogous quantum-gravity effect could be distinguished
by its different energy dependence: the quantum-gravity effect
would {\rm increase} with energy, whereas conventional medium effects
{\rm decrease} with energy in the range of interest~\cite{latorre}.

Also relevant for
building some quantum-gravity intuition for this type
of {\it in vacuo} dispersion and deformed velocity law
is the observation~\cite{aemn1,gacxt} that this has
implications for the measurability of distances in quantum gravity
that fit well with the intuition emerging from heuristic
analyses~\cite{gacmpla} based on a combination of arguments from
ordinary quantum mechanics and general relativity.
[This connection between dispersion relations and 
measurability bounds will be here reviewed in Section~8.]

Notably, recent work~\cite{gampul} has provided evidence
for the possibility that the popular
canonical/loop quantum gravity~\cite{canoloop}
might be among the theoretical approaches
that admit the phenomenon of deformed dispersion
relations with the deformation going linearly with
the Planck length ($L_p \sim 1/E_p$).
Similarly, evidence for this type of dispersion relations
has been found~\cite{aemn1} in {\it Liouville} (non-critical)
strings~\cite{emn}, whose development was partly motivated 
by an intuition concerning the ``quantum-gravity vacuum''
that is rather close to the one traditionally associated
to the works of Wheeler~\cite{wheely} and Hawking~\cite{hawk}.
Moreover,
the phenomenon of deformed dispersion
relations with the deformation going linearly with
the Planck length fits rather naturally
within certain approaches based on non-commutative geometry and 
deformed symmetries.
In particular, there is growing
evidence~\cite{gacxt,lukipapers,gacmaj}
for this phenomenon in theories living in
the non-commutative Minkowski space-time proposed
in Refs.~\cite{firstkappa,shahnkappamin,kpoin},
which involves a dimensionful (presumably Planck-length related)
deformation parameter.

Equation (\ref{vdef}) encodes a minute modification
for most practical
purposes, since $E_{QG}$ is believed to be a very high scale,
presumably of order the Planck scale $E_{p}$.
Nevertheless,
such a deformation could be rather significant for even
moderate-energy signals, if they travel over very long
distances.
According to (\ref{vdef}) two signals respectively
of energy $E$ and $E + \Delta E$
emitted simultaneously from the same
astrophysical source in traveling a distance $L$
acquire a ``relative time delay'' $|\delta t|$
given by 
\begin{eqnarray}
 |\delta t| \sim  {\Delta E \over E_{QG}} \frac{L}{c}  ~.
\label{delayt}
\end{eqnarray}
Such a time delay can be observable if $\Delta E$
and $L$ are large while
the time scale over which the signal exhibits time structure
is small.
As mentioned,
these are the respects in which GRBs offer particularly good
prospects for such measurements.
Typical photon energies in GRB
emissions are in the range $0.1-100$~MeV~\cite{review}, and it is
possible that the spectrum might in fact extend up to TeV
energies~\cite{TeV}. Moreover, time structure down to the millisecond
scale has been observed in the light curves~\cite{review}, as is
predicted in the most popular theoretical models~\cite{threv}
involving merging neutron stars or black holes, where the last stages
occur on the time scales associated with grazing orbits. Similar time
scales could also occur in models that identify GRBs with other
cataclysmic stellar events such as failed supernovae Ib, young
ultra-magnetized pulsars or the sudden deaths of massive
stars~\cite{threv2}. We see from equations (\ref{vdef}) and
(\ref{delayt}) that a signal with millisecond time structure in
photons of energy around 10~MeV coming from a distance of order
$10^{10}$ light years, which is well within the range of GRB
observations and models, would be sensitive to $E_{QG}$ of order
the Planck scale.

In order to set a definite bound on $E_{QG}$ it is necessary to
measure $L$ and to measure the time of arrival
of different energy/wavelength components of a
sharp peak within the burst.
From Eq.~(\ref{delayt}) it follows that
one could set a bound
\begin{eqnarray}
 E_{QG} > \Delta E \,  \frac{L}{c \, |\tau|}  ~
\label{ebound}
\end{eqnarray}
by establishing the times of arrival of the peak
to be the same up to an uncertainty $\tau$
in two energy channels $E$ and $E + \Delta E$.
Unfortunately, at present we have
data available only on a few GRBs for which
the distance $L$ has been determined (the first measurements
of this type were obtained only in 1997), and these are the only
GRBs which can be reliably used to set bounds on the new effect.
Moreover, mostly because of the nature of the relevant
experiments (particularly the BATSE
detector on the Compton Gamma Ray Observatory),
for a large majority of the GRBs on record only
the portion of the burst with energies up to 
the MeV energy scale was observed, whereas
higher energies would be helpful for the study of
the phenomenon of quantum-gravity-induced dispersion
here considered (which increases linearly with energy).
We expect significant improvements in these coming years.
The number of observed GRBs with associated distance
measurement should rapidly increase.
A new generation of orbiting spectrometers, {\it e.g.}
AMS~\cite{AMS} and GLAST~\cite{GLAST}, are being developed, whose
potential sensitivities are very impressive. For example, assuming a
$E^{-2}$ energy spectrum, GLAST would expect to observe about 25 GRBs
per year at photon energies exceeding 100~GeV, with time resolution of
microseconds. AMS would observe a similar number at $E>10$~GeV
with time resolution below 100~nanoseconds.

While we wait for these new experiments,
preliminary bounds can already be set with available data.
Some of these bounds are ``conditional'' in the sense that
they rely on the assumption that the relevant GRB
originated at distances corresponding to
redshift of ${\cal O}(1)$ (corresponding to a
distance of $\sim3000$~Mpc), which appears to be typical.
Let me start by considering the ``conditional'' bound
(first considered in Ref.~\cite{gacgrb})
which can be obtained from data on GRB920229.
GRB920229 exhibited~\cite{microburst}
micro-structure in its burst at energies up to $\sim200$~KeV.
In Ref.~\cite{gacgrb} it was estimated conservatively
that a detailed time-series analysis might reveal
coincidences in different BATSE energy bands on a
time-scale $\sim10^{-2}$~s, which,
assuming redshift of ${\cal O}(1)$
(the redshift of GRB920229
was not measured) 
would yield sensitivity
to $E_{QG}\sim10^{16}$~GeV (it would allow to set
a bound $E_{QG} > 10^{16}$~GeV).

As observed in Ref.~\cite{aemns2},
a similar sensitivity might be obtainable
with GRB980425, given its likely identification with the unusual
supernova 1998bw \cite{sngrb}. This is known to have taken place
at a redshift $z=0.0083$ corresponding to a
distance $D\sim40$~Mpc (for a Hubble constant
of 65 km\,sec$^{-1}$Mpc$^{-1}$) which is rather
smaller than a typical GRB distance. However GRB980425 was observed
by BeppoSAX at energies up to $1.8$~MeV, which gains back an order of
magnitude in the sensitivity. If a time-series analysis
were to reveal structure at
the $\Delta\,t\sim10^{-3}$~s level, which
is typical of many GRBs~\cite{variation}, it would yield the same
sensitivity as GRB920229 (but in this case,
in which a redshift measurement is available,
one would have a definite bound, whereas GRB920229
only provides a ``conditional''
bound of the type discussed above).

Ref.~\cite{aemns2} also observed that an interesting
(although not very ``robust'')
bound could be obtained using GRB920925c,
which was observed by WATCH~\cite{watch}
and possibly in high-energy $\gamma$ rays by the
HEGRA/AIROBICC array above $20$~TeV~\cite{hegra}.
Several caveats are
in order: taking into account the appropriate trial factor, the
confidence level for the signal seen by HEGRA to be related to
GRB920925c is only $99.7\%$ ($\sim2.7\sigma$), the reported
directions differ by 9$^0$, and the redshift of the source is
unknown. Nevertheless, the potential sensitivity is impressive. The
events reported by HEGRA range up to $ E\sim200$~TeV, and the
correlation with GRB920925c is within $\Delta\,t\sim200$~s. Making
the reasonably conservative assumption that GRB920925c occurred no
closer than GRB980425, viz. $\sim40$Mpc, one finds a minimum
sensitivity to $E_{QG}\sim10^{19}$~GeV,
modulo the caveats listed above. Even more
spectacularly, several of the HEGRA GRB920925c candidate
events occurred within $\Delta\,t\sim1$~s, providing a
potential sensitivity even two orders of magnitude higher.

As illustrated by this discussion, the GRBs have remarkable potential
for the study of {\it in vacuo} dispersion, which will certainly
lead to impressive bounds/tests as soon as improved experiments
are put into operation, but at present the best GRB-based bounds
are either ``conditional'' (example of GRB92022)
or ``not very robust''(example of GRB920925c).
As a result, at present the best (reliable) bound has been
extracted~\cite{billetal,Whipple}
using data from the Whipple telescope on a
TeV $\gamma$-ray flare associated with the active galaxy Mrk~421.
This object has a redshift of 0.03 corresponding to a distance of
$\sim100$~Mpc. Four events with $\gamma$-ray energies above 2~TeV have
been observed within a period of $280$~s. These provide~\cite{billetal,Whipple}
a definite limit $E_{QG}>4\times10^{16}$~GeV.

In passing let me mention that (as observed
in Ref.~\cite{gacgrb,aemn1})
pulsars and
supernovae, which are among
the other astrophysical phenomena that might at
first sight appear well suited 
for the study of {\it in vacuo} dispersion, 
do not actually provide interesting sensitivities.
Although pulsar signals have very well-defined time
structure, they are at relatively low energies and are
presently
observable over distances of at most $10^4$ light years.
If one takes an energy of
order $1$ eV and postulates generously a sensitivity to time delays
as small as $1~\mu$sec, one nevertheless reaches only
an estimated sensitivity to $E_{QG} \sim 10^{9}$~GeV.
With new experiments such as AXAF it may be possible
to detect X-ray pulsars out to $10^6$ light years,
but this would at best push the sensitivity up
to $E_{QG} \sim 10^{11}$~GeV.
Concerning supernovae,
it is important to take into account that
neutrinos\footnote{Of course, at present
we should be open to the possibility that the
velocity law (\ref{vdef}) might apply to all
massless particles, but it is also plausible that
the correct quantum-gravity velocity law would depend
on the spin of the particle. It would therefore be
important to set up a phenomenological programme
that studies neutrinos with the same level of sensitivies
that GRBs and other astrophysical phenomena allow for
the study of the velocity law of the photon.}
from Type II events
similar to SN1987a, which should have energies up to about 100~MeV
with a time structure that could extend down to milliseconds, are
likely to be detectable at distances of up to about $10^{5}$ light
years, providing sensitivity to $E_{QG} \sim 10^{15}$~GeV,
which is remarkable in absolute terms, but is still significantly
far from the Planck scale and anyway cannot compete with
the type of sensitivity achievable with GRBs.

It is rather amusing that GRBs can provide such
a good laboratory for
investigations of {\it in vacuo} dispersion
in spite of the fact that
the short-time structure of GRB signals
is still not understood.
The key point of the proposal
in Ref.~\cite{gacgrb} is that sensitive tests can be
performed through the serendipitous detection of
short-scale time
structure~\cite{microburst} at different energies in
GRBs which are established to be at cosmological distances.
Detailed features of burst time series enable
(as already shown in several examples) the
emission times in different energy ranges to be put into
correspondence.
Any time shift due to quantum-gravity
would {\it increase} with the photon energy, and
this characteristic dependence is separable from more
conventional in-medium-physics effects, which {\it decrease}
with energy. To distinguish any quantum-gravity
induced shift from effects due to the source,
one can use the fact that the
quantum-gravity effect here considered is {\it linear} in
the GRB distance.

This last remark applies to all values of $\alpha$,
but most of the observations and formulas  
in this section are only valid in the case $\alpha = 1$
(linear suppression).
The generalization to cases with $\alpha \ne 1$
is however rather simple; for example, Eq.~(\ref{ebound})
takes the form (up to coefficients of order 1)
\begin{eqnarray}
 E_{QG} > \left[ \left[ (E+\Delta E)^\alpha - E^\alpha \right]
\frac{L}{c \, |\tau|} \right]^{1/\alpha}  ~.
\label{eboundgen}
\end{eqnarray}
Notice that here, because of the non-linearity,
the right-hand side depends both on $E$ and $\Delta E$.

Before moving on to other experiments let me
clarify what is the key ingredient of this experiment using
observations of gamma
rays from distant astrophysical sources
(the ingredient that allowed to
render observable the minute quantum-gravity effects).
This ingredient is very similar to the one relevant for
the studies of space-time
fuzziness using modern interferometers which I discussed in
the preceding section; in fact,
the gamma rays here considered are affected by
a very large number of the minute quantum-gravity effects.
Each of the dispersion-inducing quantum-gravity effect
is very small, but the gamma rays
emitted by distant astrophysical sources
travel for a very long time before reaching us
and can therefore be affected
by an extremely large number of such effects.

\section{OTHER QUANTUM-GRAVITY EXPERIMENTS}

In this section I provide brief reviews of 
some other quantum-gravity experiments.
The fact that 
the discussion here provided for these experiments
is less detailed than the preceding discussions
of the interferometry-based and GRB-based experiments
is not to be interpreted as indicating that these experiments
are somehow less significant: it is just that
a detailed discussion of a couple 
of examples was sufficient to provide to the reader 
some general intuition on the strategy behind quantum-gravity
experiments and it was natural for me to 
use as examples the ones I am most familiar with.
For the experiments discussed in this section I shall just 
give a rough idea of the quantum-gravity scenarios that
could be tested and of the experimental procedures which have
been proposed.

\subsection{Neutral kaons and CPT violation}

One of the formalisms that has been proposed~\cite{ehns,elmn} 
for the evolution equations of particles in the space-time foam
relies on a density-matrix picture.
The foam is seen as providing a sort of environment
inducing quantum decoherence even on isolated systems
({\it i.e.} systems which only interact with the foam).
A given non-relativistic system (such as the neutral kaons 
studied by the CPLEAR collaboration at CERN)
is described by a density matrix $\rho$
that satisfies an evolution equation analogous to the
one ordinarily used for the quantum mechanics of 
certain open systems:
\begin{eqnarray}
 \partial_t \rho = i [\rho,H] + \delta \! H \, \rho
\label{ehnseq}
\end{eqnarray}
where $H$ is the ordinary Hamiltonian and $\delta \! H$, which has
a non-commutator structure~\cite{elmn}, represents the effects of the foam.
$\delta \! H$ is expected to be extremely small, suppressed by some
power of the Planck length.
The precise form of $\delta \! H$ (which in particular would
set the level of the new physics by establishing how many powers
of the Planck length suppress the effect)
has not yet been derived from some full-grown quantum 
gravity\footnote{Within the 
quantum-gravity approach here reviewed in Subsection 11.2,
which only attempts to model certain aspects of quantum gravity,
such a direct calculation might soon be performed.},
and therefore phenomenological parametrizations have been
introduced (see Refs.~\cite{ehns,elmncptheory,hpcpt,floreacpt}).
For the case in which the effects are only suppressed
by one power of the Planck length (linear suppression)
recent neutral-kaon experiments, such as the ones
performed by CPLEAR,
have set significant bounds~\cite{elmn} on the
associated CPT-violation effects and forthcoming experiments
are likely to push these bounds even further.

Like the interferometry-based and the GRB-based experiments,
these experiments (which have the added merit of having started the
recent wave of quantum-gravity proposals)
also appear to provide significant quantum-gravity tests.
As mentioned, the effect of quantum-gravity-induced decoherence 
certainly qualifies as a traditional quantum-gravity
subject, and the level of sensitivity reached by the
neutral-kaon studies is certainly significant (as in the case 
of  {\it in vacuo} dispersion and GRBs, one would like to
be able to explore also the case of a quadratic Planck-length
suppression, but it is nonetheless very significant
that we have
at least reached the capability to test the case of linear
suppression).
Also in this case it is natural to ask: how come we could manage this?
What strategy allowed this neutral-kaon studies to evade
the traditional gloomy forecasts for quantum-gravity phenomenology?
While, as discussed above, 
in the interferometry-based and the GRB-based experiments
the crucial element in the experimental
proposal is the possibility to put together many quantum gravity
effects, in the case of the neutral-kaon experiments the crucial
element in the experimental
proposal is provided by the very delicate balance of scales that
characterizes the neutral-kaon system. In particular,
it just happens to be true that
the dimensionless ratio setting the 
order of magnitude of quantum-gravity
effects in the linear suppression scenario,
which is $c^2 M_{L,S}/E_p \sim 2 \cdot 10^{-19}$,
is not much smaller than  
some of the dimensionless ratios characterizing the
neutral-kaon system, notably the 
ratio $|M_L - M_S|/M_{L,S} \sim 7 \cdot 10^{-15}$
and the ratio $|\Gamma_L - \Gamma_S|/M_{L,S} \sim 1.4 \cdot 10^{-14}$.
This renders possible for the
quantum-gravity effects to provide observably large corrections
to the physics of neutral kaons.

\subsection{Interferometry and string cosmology}

Up to this point I have only reviewed experiments probing
foamy properties of space-time in the sense of Wheeler and Hawking.
A different type of quantum-gravity effect which might
produce a signature strong enough for experimental testing
has been discussed in the context of studies 
of a cosmology based on
critical superstrings~\cite{stringcosm}.
While for a description of this cosmology and 
of its physical signatures I must only refer the reader
to the recent reviews in Ref.~\cite{stringcoreviews},
I want to briefly discuss here the basic ingredients
of the proposal~\cite{stringcogwi} of interferometry-based tests
of the stochastic gravity-wave background predicted
by string cosmology.

In string cosmology the universe starts from a 
state of very small curvature, then goes through a
long phase of dilaton-driven inflation reaching nearly Plankian
energy density, and then eventually reaches the standard 
radiation-dominated 
cosmological evolution~\cite{stringcosm,stringcoreviews}.
The period of nearly Plankian
energy density plays a crucial role in rendering the
quantum-gravity effects observable. In fact, this
example based on string cosmology is quite different from
the experiments I discussed earlier in these lectures
also because it does not involve
small quantum-gravity effects which are somehow amplified 
(in the sense for example of the amplification provided when
many effects are somehow put together).
The string cosmology involves a period in which the quantum-gravity
effects are actually quite large.
As clarified in Refs.~\cite{stringcosm,stringcoreviews}
planned gravity-wave detectors such as LIGO might be able to detect
the faint residual traces of these strong effects occurred in
a far past.

As mentioned, the quantum-gravity effects that,
within string cosmology, leave a trace in
the gravity-wave background are not of the type
that requires an active Wheeler-Hawking foam.
The relevant quantum-gravity effects live in 
the more familiar vacuum which we are used to
encounter in the context of ordinary gauge theory.
(Actually, for the purposes of the analyses
reported in Refs.~\cite{stringcosm,stringcoreviews}
quantum gravity could be seen as an ordinary gauge theory,
although with unusual gauge-field content.)
In the case of the Wheeler-Hawking foam one is tempted to
visualize the vacuum 
as reboiling with (virtual) worm-holes and black-holes.
Instead the effects relevant for the gravity-wave background
predicted by string cosmology
are more conventional field-theory-type fluctuations,
although carrying gravitational degrees of freedom, like the graviton.
Also from this point of view the experimental proposal
discussed in Refs.~\cite{stringcosm,stringcoreviews}
probes a set of candidate quantum-gravity phenomena
which is complementary to the ones I have reviewed earlier in these
notes.

\subsection{Matter interferometry and primary state diffusion}

The studies reported in Ref.~\cite{peri} (and references
therein) have considered how certain effectively
stochastic properties of space-time would affect
the evolution of quantum-mechanical states.
The stochastic properties there considered are different from
the ones discussed here in Sections 2, 3 and 4, but were introduced
within a similar viewpoint, {\it i.e.} stochastic processes
as effective description of quantum space-time processes.
The implications of these stochastic properties
for the evolution of quantum-mechanical states
were modeled via the formalism of ``primary  state diffusion'',
but only rather crude models turned out to be treatable.

The approach proposed in Ref.~\cite{peri} actually puts together
some of the unknowns of space-time foam and the specific 
properties of ``primary  state diffusion''. The structure of
the predicted effects cannot be simply discussed in terms of 
elementary properties of space-time foam and
a simple interpretation in terms of symmetry deformations
does not appear to be possible.
Those effects
appear to be the net result of the whole formalism that goes into 
the approach. Moreover, as also emphasized by the authors, the 
crudeness of the models is such that all conclusions are to be 
considered as tentative at best. Still, the analysis
reported in Ref.~\cite{peri} is very significant
as an independent indication of a mechanism,
based on matter-interferometry experiments, that could unveil
Planck-length-suppressed effects.

\subsection{Colliders and large extra dimensions}

It was recently suggested~\cite{anto,addlarge}
that the characteristic quantum-gravity length scale might be 
given by a length scale $L_D$ much larger
than the Planck length
in theories with large extra dimensions.
It appears plausible that there exist models that are
consistent with presently-available experimental data
and have $L_D$ as large as 
the $(TeV)^{-1}$ scale and (some of) the extra dimensions 
as large as a millimiter~\cite{addlarge}.
In such models the smallness of the Planck length
is seen as the result of the fact that the strength
of gravitation in the ordinary 3+1 space-time dimensions 
would be proportional to the square-root of the inverse of
the large volume of the external compactified space multiplied
by an appropriate (according to dimensional analysis)
power of $L_D$.

Several studies have been motivated by
the proposal put forward in Ref.~\cite{addlarge},
but only a small 
percentage of these studies
considered the implications for quantum-gravity scenarios.
Among these studies the ones reported in Refs.~\cite{grwlarge,grwlarge2}
are particularly significant for the objectives of these lectures, 
since they illustrate another completely different strategy
for quantum-gravity experiments.
It is there observed that within the realm 
of the ordinary 3+1 dimensional space-time
an important consequence of the existence 
of large 
extra dimensions would be the presence
of a tower of Kaluza-Klein modes associated to the gravitons.
The weakness of the coupling between gravitons and other particles 
can be compensated by the large number of these Kaluza-Klein modes
when the experimental energy resolution
is much larger than the mass splitting between the modes,
which for a small number of 
very large extra dimensions can be a weak requirement 
({\it e.g.} for 6 millimiter-wide extra
dimensions~\cite{addlarge,grwlarge}
the mass splitting is of a few $MeV$).
This can lead to observably large~\cite{grwlarge,grwlarge2} effects at
planned particle-physics colliders, particularly CERN's LHC.

In a sense, the 
experimental proposal put forward in Refs.~\cite{grwlarge,grwlarge2}
is another example of experiment
in which the smallness of quantum gravity effects 
is compensated by putting together a large number of such effects
(putting together the contributions of all of the Kaluza-Klein modes).

Concerning the quantum-gravity aspects of the 
models with large extra dimensions
proposed in Ref.~\cite{addlarge}, it is important to realize that,
as emphasized in Ref.~\cite{bignapap}, if anything like the space-time
foam here described in Sections 2, 3, 4 and 5 was present in such models
the effective reduction of the quantum-gravity scale
would naturally lead to foamy effects that are too large
for consistency with available experimental data.
Preliminary estimates based solely on dimensional considerations
appear to suggest that~\cite{bignapap} 
linear suppression
by the reduced quantum-gravity scale would certainly be ruled out 
and even quadratic
suppression might not be sufficient for consistency with available data.
These arguments should lead to rather stringent
bounds on space-time foam especially in those models in which
some of the large extra dimensions are accessible to non-gravitational 
particles (see, {\it e.g.}, Ref.~\cite{photolarge}),
and should have interesting (although smaller)
implications also for
the popular scenario in which only the gravitational degrees of freedom
have access to the large extra dimensions.
Of course, a final verdict must await detailed calculations
analysing space-time foam in these models with large extra
dimensions. 
The first examples of
this type of computations
are given by the very recent studies
in Refs.~\cite{fordlarge,adrian}, which
considered the implications of foam-induced light-cone deformation
for certain examples of models with large extra dimensions.

\section{CLASSICAL-SPACE-TIME-INDUCED QUANTUM PHASES 
IN MATTER INTERFEROMETRY}

While of course the quantum properties of space-time 
are the most exciting effects we expect of quantum gravity,
and probably the ones which will prove most useful
in gaining insight into the fundamental
structure of the theory,
it is important to investigate experimentally
all aspects of the interplay between
gravitation and quantum mechanics.
Among these experiments the ones that could be expected to 
provide fewer surprises (and less insight into the structure
of quantum gravity)
are the ones concerning the interplay between strong-but-classical
gravitational fields and quantum matter fields.
However, this is not necessarily true
as I shall try to clarify within this section's brief
review of the experiment performed nearly a quarter of a 
century ago by Colella, Overhauser and Werner~\cite{cow},
which, to my knowledge, was the first experiment
probing some aspect of the interplay between
gravitation and quantum mechanics.
That experiment has been followed by several modifications
and refinements (often labeled ``COW experiments'' from the initials
of the scientists involved in the first experiment)
all probing the same basic physics, {\it i.e.} the validity
of the Schr\"odinger equation
\begin{eqnarray}
\left[ - \left( {\hbar^2 \over 2 \, M_I} \right) \vec{\nabla}^2 
+ M_G \, \phi(\vec{r}) \right] \psi(t,\vec{r}) = i \, \hbar \,
{\partial \, \psi(t,\vec{r}) \over \partial t}
\label{coweq}
\end{eqnarray}
for the description of the dynamics of matter (with wave 
function $\psi(t,\vec{r})$) in presence of the Earth's 
gravitational potential $\phi(\vec{r})$.
[In (\ref{coweq}) $M_I$ and $M_G$ denote the inertial
and gravitational mass respectively.]

The COW experiments exploit the fact that
the Earth's 
gravitational potential puts together the contribution
of so many particles (all of those composing the Earth)
that it ends up being large enough to introduce observable
effects in rotating table-top interferometers.
This was the first example of a physical context in which
gravitation was shown to have an observable effect on
a quantum-mechanical system in spite of 
the weakness of the gravitational force.

The fact that the original experiment performed 
by Colella, Overhauser and Werner
obtained results in
very good agreement~\cite{cow} with
Eq.~(\ref{coweq}) might seem to indicate
that, as generally expected, experiments on
the interplay between strong-but-classical
gravitational fields and quantum matter fields
should not lead to surprises and should not provide
insight into the structure of quantum gravity.
However, the confirmation of Eq.~(\ref{coweq})
does raise some sort of a puzzle with respect to 
the Equivalence Principle of general relativity;
in fact, even for $M_I=M_G$ 
the mass does not cancel out in 
the quantum evolution equation (\ref{coweq}).
This is an observation that by now has also been emphasized in
textbooks~\cite{textbook}, but to my knowledge it has not been
fully addressed even within the most popular
quantum-gravity approaches, {\it i.e.} critical
superstrings
and canonical/loop quantum gravity.
Which role should be played by the Equivalence Principle in 
quantum gravity?
Which version/formulation of the Equivalence Principle 
should/could hold in quantum gravity?

Additional elements for consideration in quantum-gravity
models will emerge if the small discrepancy
between (\ref{coweq}) and data 
reported in Ref.~\cite{wernb}
(a refined COW experiment)
is confirmed by future experiments.
The subject of gravitationally
induced quantum phases
is also expanding in new
directions~\cite{ahluexp,dharamnature},
which are likely to provide additional insight.

\section{ESTIMATES OF SPACE-TIME FUZZINESS
FROM MEASURABILITY BOUNDS}

In the preceding Sections 4, 5, 6 and 7
I have discussed the experimental proposals that
support the conclusions anticipated in Sections 2 and 3.
This Section 8 and the following three sections
each provide a ``theoretical-physics addendum''.
In this section I discuss some arguments
that appear to suggest properties of the space-time foam.
These arguments are based on analyses of bounds on the
measurability of distances in quantum gravity.
The existence of measurability
bounds has attracted the interest of several
theorists, because these bounds
appear to capture an important novel element
of quantum gravity. In ordinary (non-gravitational)
quantum mechanics there is no absolute limit on the
accuracy of the measurement of a distance.
[Ordinary quantum mechanics allows $\delta A = 0$ for any
single observable $A$, since it only limits the combined
measurability of pairs of conjugate observables.]

The quantum-gravity bound on the measurability of distances
(whatever final form it
actually takes in the correct theory)
is of course intrinsically interesting,
but here (as in previous
works \cite{gacgwi,bignapap,gacmpla,areapap,gacgrf98})
I shall be interested in the possibility that 
it might reflect properties of the space-time foam.
This is of course not necessarily true:
a bound on the measurability of distances is
not necessarily associated to space-time fluctuations,
but guided by the Wheeler-Hawking intuition on the
nature of space-time one is tempted to interpret any measurability
bound (which might be obtained with totally independent arguments)
as an indicator of the type of irreducible fuzziness
that characterizes space-time.
One has on one hand some intuition about quantum gravity
which involves stochastic fluctuations of distances
and on the other hand some different arguments lead to intuition
for absolute bounds on the measurability of distances;
it is natural to explore the possibility that the two
might be related, {\it i.e.} that the intrinsic stochastic
fluctuations should limit the measurability just
to the level suggested by
the independent measurability arguments.
Different arguments appear to lead to different
measurability bounds, which in turn could provide
different intuition for the stochastic properties of
space-time foam.

\subsection{Minimum-length noise}

In many quantum-gravity approaches there appears to be
a length scale $L_{min}$, often identified with 
the Planck length
or the string length $L_{string}$ (which, as mentioned, should
be somewhat larger than the Planck length, plausibly
in the neighborhood of $10^{-34}m$),
which sets an absolute bound
on the measurability of distances (a minimum uncertainty):
\begin{eqnarray}
\delta D \ge L_{min}
~. \label{minlen}
\end{eqnarray}
This property emerges in approaches based on canonical
quantization of Einstein's gravity when analyzing
certain gedanken experiments
(see, {\it e.g.}, Refs.~\cite{padma,garay} and references therein).
In critical superstring theories, theories whose mechanics is 
still governed
by the laws of ordinary quantum mechanics but with one-dimensional
(rather than point-like) fundamental objects,
a relation of type (\ref{minlen}) follows from
the stringy modification
of Heisenberg's uncertainty principle~\cite{venekonmen}
\begin{eqnarray}
 \delta x \, \delta p \!\!& \ge &\!\! 1
+ {L_{string}^2} \, \delta p^2
~. \label{veneup}
\end{eqnarray}
In fact, whereas Heisenberg's uncertainty principle
allows $\delta x = 0$ (for $\delta p \rightarrow \infty$),
for all choices of  $\delta p$ 
the uncertainty relation (\ref{veneup})
gives $\delta x \ge L_{string}$.
The relation (\ref{veneup}) is suggested by certain analyses
of string scattering~\cite{venekonmen}, but it
might have to be modified when taking into account
the non-perturbative solitonic structures of superstrings
known as Dirichlet branes~\cite{dbrane}.
In particular, evidence has been found~\cite{dbrscatt}
in support of the possibility that ``Dirichlet
particles" (Dirichlet 0~branes)
could probe the structure of space-time down
to scales shorter than the string length.
In any case, all evidence available on critical superstrings
is consistent with a relation of type (\ref{minlen}),
although it is probably safe to say that
some more work is still needed to firmly establish 
the string-theory value of $L_{min}$.

Having clarified that a relation of type (\ref{minlen})
is a rather common prediction of theoretical work
on quantum gravity,
it is then natural 
to wonder whether such a relation
is suggestive of stochastic distance fluctuations
of a type that 
could significantly affect the noise levels of an interferometer.
As mentioned, relations such as (\ref{minlen})
do not necessarily encode any fuzziness; for example,
relation (\ref{minlen}) could simply emerge from
a theory based on a lattice of points with spacing $L_{min}$
and equipped with a measurement theory consistent
with (\ref{minlen}).
The concept of distance in such a theory
would not necessarily be affected by the
type of stochastic processes that lead to noise in
an interferometer.
However,
if one does take as guidance the Wheeler-Hawking intuition
on space-time foam it makes sense to assume that
relation (\ref{minlen}) might
encode the net effect of some underlying physical processes
of the type one would qualify as quantum space-time fluctuations.
This (however preliminary) network of intuitions suggests that
(\ref{minlen}) could be the result of fuzziness for distances $D$
of the type associated with stochastic fluctuations with
root-mean-square deviation $\sigma_D$ given by
\begin{equation}
\sigma_D \sim L_{min} \, .
\label{no1}
\end{equation}
The associated
displacement amplitude spectral density $S_{min}(f)$
should roughly have a $1/\sqrt{f}$ behaviour
\begin{equation}
S_{min}(f) \sim {L_{min} \over \sqrt{f}} \, ,
\label{no1spectrum}
\end{equation}
which (using notation set up in Section~4)
can be concisely described stating
that $L_{min} \sim {\cal L}_{\beta=1/2}$.
Eq.~(\ref{no1spectrum}) can be justified using the
general relation (\ref{gacspectrule}).
Substituting the $S_{min}(f)$ of Eq.~(\ref{no1spectrum})
for the $S(f)$ of Eq.~(\ref{gacspectrule})
one obtains a $\sigma$ that approximates the $\sigma_D$
of Eq.~(\ref{no1}) up to small (logarithmic) $T_{obs}$-dependent
corrections.
A more detailed description of the
displacement amplitude spectral density associated
with Eq.~(\ref{no1}) can be found in Refs.~\cite{jare1,jare2}.
For the objectives of these lectures the rough
estimate (\ref{no1spectrum}) is sufficient since,
if indeed $L_{min} \sim L_{p}$, from (\ref{no1spectrum}) 
one obtains $S_{min}(f) \sim 10^{-35} m / \sqrt{f}$,
which is still very far from the sensitivity of even the most 
advanced modern
interferometers, and therefore I shall not be concerned with
corrections to Eq.~(\ref{no1spectrum}).

\subsection{Random-walk noise motivated by the analysis of
a Salecker-Wigner gedanken experiment}

Let me now consider a measurability bound which is encountered
when taking into account the quantum properties of devices.
It is well understood (see, {\it e.g.},
Refs.~\cite{gacmpla,gacgrf98,bergstac,diosi,ng,dharam94grf}) 
that the combination of 
the gravitational properties 
and 
the quantum
properties 
of devices can have an important
role in the analysis
of the operative definition of gravitational observables.
Since the analyses~\cite{padma,garay,venekonmen,dbrscatt}
that led to the proposal of Eq.~(\ref{minlen}) 
only treated the devices in a completely idealized manner
(assuming that one could ignore any contribution to
the uncertainty in the measurement of $D$ due to the
gravitational and quantum properties of devices),
it is not surprising that analyses taking into account the 
gravitational and quantum properties of devices found more 
significant limitations to the measurability of distances.

Actually, by ignoring the way in which the gravitational properties 
and the quantum properties of devices combine in measurements
of geometry-related physical properties of a system
one misses some of the fundamental elements
of novelty we should expect for the interplay of gravitation
and quantum mechanics; in fact, one would be missing an
element of novelty which is deeply associated to the Equivalence
Principle.
In measurements of physical properties
which are not geometry-related one can safely
resort to an idealized description of devices.
For example, in
the famous Bohr-Rosenfeld analysis~\cite{rose}
of the measurability of the electromagnetic field
it was shown that the accuracy allowed by
the formalism of ordinary quantum mechanics could only be achieved
using idealized test particles with vanishing ratio between
electric charge and inertial mass.
Attempts to generalize the Bohr-Rosenfeld analysis
to the study of gravitational fields
(see, {\it e.g.}, Ref.~\cite{bergstac})
are of course confronted with the fact that
the ratio between gravitational ``charge'' (mass) and inertial mass
is fixed by the Equivalence Principle.
While ideal devices with vanishing ratio between  
electric charge and inertial mass can
be considered at least in principle,
devices with vanishing ratio between  
gravitational mass and inertial mass 
are not admissible in any (however formal) limit
of the laws of gravitation.
This observation provides one of the strongest elements
in support of the idea~\cite{gacgrf98}
that the mechanics on which quantum
gravity is based must not be exactly
the one of ordinary quantum mechanics, since it should
accommodate a somewhat different relationship between ``system''
and ``measuring apparatus''
and should not rely on the idealized ``measuring apparatus''
which plays such a central role in the mechanics laws of
ordinary quantum mechanics (see, {\it e.g.}, 
the ``Copenhagen interpretation'').

In trying to develop some intuition for the type
of fuzziness that could affect the concept of distance
in quantum gravity, it might be useful to consider the way 
in which the interplay between the
gravitational and the quantum properties of devices affects
the measurability of distances.
In Refs.~\cite{gacmpla,gacgrf98} I have argued\footnote{I
shall comment later in these notes on the measurability
analysis reported
in Ref.~\cite{ng}, which also took
as starting point the mentioned
work by Salecker and Wigner, but
advocated a different viewpoint and reached
different conclusions.}
that a natural starting point for this type of analysis
is provided by the procedure for the measurement of distances
which was discussed in 
influential work\footnote{The classic Salecker-Wigner
work~\cite{wign} is criticized in the
recent paper~\cite{anos}.
As I explain in detail in Ref.~\cite{swok},
the analysis reported in Ref.~\cite{anos}
is incorrect.
Whereas Salecker and Wigner sought an operative definition
of distances suitable for the Planck regime,
the analysis in Ref.~\cite{anos}
relies on several assumptions
that appear to be natural in the context of
most present-day experiments
but are not even meaningful in the Planck regime.
Moreover, contrary to the claim made in Ref.~\cite{anos},
the source of $\sqrt{T_{obs}}$-uncertainty
used in the Salecker-Wigner derivation cannot be 
truly eliminated;
unsurprisingly, it can only be traded~\cite{swok}
for another comparable 
contribution to the total uncertainty in the measurement.
In addition to this incorrect criticism
of the limit derived by Salecker and Wigner,
Ref.~\cite{anos} also misrepresented the role
of the Salecker-Wigner
limit in providing motivation
for the interferometric studies here considered
(and originally proposed in Refs.~\cite{gacgwi,bignapap}):
the reader could come out of reading Ref.~\cite{anos}
with the impression that such interferometry-based
tests would only be sensitive to quantum-gravity ideas
motivated by the Salecker-Wigner limit.
As emphasized in Sections 4 and 8 of
these notes (and in Ref.~\cite{bignapap})
motivation for this phenomenological programme 
also comes from a long tradition of ideas
(developing independently of the ideas related to
the Salecker-Wigner limit)
on foamy/fuzzy space-time,
and from recent work
on the possibility that quantum-gravity
might induce a deformation of the dispersion relation that
characterizes the propagation of the massless particles
used as space-time probes in the operative definition
of distances.
This is already quite clear at least to a portion of the community;
for example, in recent work~\cite{adrian} on foamy space-times
(without any reference to the Salecker-Wigner related literature)
the type of modern-interferometer sensitivity exposed 
in Refs.~\cite{gacgwi,bignapap}
was used
to constrain certain novel candidate quantum-gravity effects.}
by Salecker and Wigner~\cite{wign}.
These authors ``measured'' (in the ``{\it gedanken}'' sense)
the distance $D$ between two bodies 
by exchanging a light signal between them.
The measurement procedure requires {\it attaching}\footnote{Of
course, for consistency with causality,
in such contexts one assumes devices to be ``attached non-rigidly,''
and, in particular, the relative position
and velocity of their centers of mass continue to satisfy the
standard uncertainty relations of quantum mechanics.} 
a light-gun ({\it i.e.} a device 
capable of sending
a light signal when triggered), a detector
and a clock to
one of the two bodies 
and {\it attaching} a mirror to the other body. 
By measuring the time $T_{obs}$ (time of observation)
needed by the light signal
for a two-way journey between the bodies one 
also
obtains a 
measurement of  
the distance $D$.
For example, in flat space 
and neglecting quantum effects 
one simply finds that $D = c {T_{obs} / 2}$.
Within this setup it is easy to realize that the
interplay between the
gravitational and the quantum properties of devices 
leads to an irreducible contribution to the uncertainty $\delta D$.
In order to see this it is sufficient to consider the
contribution to $\delta D$ coming from 
the uncertainties that affect the motion
of the center of mass of the system
composed by the light-gun, the detector and the clock.
Denoting with $x^*$ and $v^*$
the position and the velocity of the center of mass
of this composite device
relative to the position of the body to which it is {\it attached},
and assuming that the experimentalists prepare this device
in a state characterised by
uncertainties $\delta x^*$ and $\delta v^*$,
one easily finds~\cite{wign,gacgrf98}
\begin{eqnarray}
\delta D \geq 
\delta x^* + T_{obs} \delta v^* 
\geq 
\delta x^* 
+ \left( {1 \over  M_b} + {1 \over  M_d} \right)
{ \hbar T_{obs} \over 2 \, \delta x^* }
\geq \sqrt{ {\hbar T_{obs} \over 2}
{1 \over  M_d} }
~,
\label{deltawignOLD}
\end{eqnarray}
where $M_b$ is the mass of 
the body, $M_d$ is the total mass of the device composed of
the light-gun, the detector, and the clock,
and
I also used the fact 
that Heisenberg's {\it Uncertainty Principle} 
implies $\delta x^* \delta v^* \ge (1/M_b + 1/M_d) \hbar/2$.
[The {\it reduced mass} $(1/M_b+1/M_d)^{-1}$ 
is relevant for the relative motion.]
Clearly, from (\ref{deltawignOLD}) it follows that 
in order to reduce the contribution to the uncertainty
coming from the quantum properties of the devices
it is necessary to take
the formal ``classical-device limit,''
{\it i.e.} the limit\footnote{A body of finite mass
can acquire a nearly-classical behaviour when
immerged in a suitable environment (environment-induced
decoherence).
However, one of the central hypothesis of the work of Salecker
and Wigner and followers is that
the quantum properties of devices should not be negligible in
quantum gravity, and that in particular
the in-principle operative
definition of distances (which we expect
to lie at the foundations of quantum gravity)
should not rely on environment-induced decoherence.
It appears worth exploring the implications of this hypothesis
not only because quantum gravity could be a truly fundamental 
theory (rather than the effective large-distance description
of a more fundamental theory) but also because the operative
definition of distances in quantum gravity should be applicable
all the way down to the Planck length.
It is even plausible~\cite{wign,rovellimrs}
that quantum gravity should
accommodate an operative definition of a material
reference system composed of a network of free-falling particles
with relative distances comparable to the Planck length.
Within the framework of these intuitions it is indeed quite hard to
imagine a decoherence-inducing 
environment suitable for the in-principle operative
definition of distances in quantum gravity.
As emphasized in Ref.~\cite{swok}, the analysis reported
in Ref.~\cite{anos} missed this important conceptual element
of the Salecker-Wigner approach.}
of infinitely large $M_d$.

Up to this point I have not yet taken into account the
gravitational properties of the devices and in fact
the ``classical-device limit'' encountered above
is fully consistent with the laws of ordinary
quantum mechanics.
From a physical/phenomenological and conceptual
viewpoint it is well understood that the formalism
of quantum mechanics is only appropriate for the
description of the results of measurements performed
by classical devices. It is therefore not surprising
that the classical-device (infinite-mass) limit turns
out to be required 
in order to match the prediction $min \delta D = 0$
of ordinary quantum mechanics.

If one also takes into account
the gravitational properties of the devices,
a conflict with ordinary quantum mechanics
immediately arises because the
classical-device (infinite-mass) limit
is in principle inadmissible for measurements
concerning gravitational effects.\footnote{This conflict
between the infinite-mass classical-device limit
(which is implicit in the applications of the formalism of
ordinary quantum mechanics to the description of the outcome
of experiments)
and the nature of gravitational interactions 
has not been addressed within any of the
most popular quantum gravity approaches,
including
critical superstrings~\cite{dbrane,critstring}
and canonical/loop
quantum gravity~\cite{canoloop}.
In a sense somewhat similar to the one appropriate for
Hawking's work on black holes~\cite{hawkbh},
this ``classical-device paradox''
appears to provide an obstruction~\cite{gacgrf98} for the use
of the ordinary formalism of quantum mechanics
for a description of quantum gravity.}
As the devices get more and more massive they increasingly 
disturb the gravitational/geometrical observables, and
well before reaching the infinite-mass limit the procedures 
for the measurement of gravitational observables cannot
be meaningfully performed~\cite{gacmpla,gacgrf98,ng}.
In the Salecker-Wigner measurement procedure
the limit $M_d \rightarrow \infty$
is not admissible when gravitational interactions
are taken into account.
At the very least the value of $M_d$ is limited by the
requirement that the apparatus should not turn into a black hole
(which would not allow the exchange of signals
required by the measurement procedure).

These observations render unavoidable 
the $\sqrt{T_{obs}}$-dependence of Eq.~(\ref{deltawignOLD}).
It is important to realize that this $\sqrt{T_{obs}}$-dependence
of the bound on the measurability of distances
comes simply from combining elements of quantum mechanics
with elements of classical gravity.
As it stands it is not to be interpreted as a quantum-gravity
effect. However, as clarified in the opening of this section,
if one is interested in modeling properties of the space-time
foam it is natural to explore the possibility that
the foam be such that distances be affected by stochastic
fluctuations with this typical $\sqrt{T_{obs}}$-dependence.
The logic is here the one of observing that
stochastic fluctuations associated to the foam
would anyway lead to some form of
dependence on $T_{obs}$ and in guessing the specific form
of this dependence the measurability analysis
reviewed in this subsection can be seen as providing motivation
for a $\sqrt{T_{obs}}$-dependence.
From this point of view 
the measurability analysis
reviewed in this subsection
provides additional motivation for the study of
random-walk-type models of distance fuzziness,
whose fundamental stochastic fluctuations
are characterized (as already discussed in Section~4)
by root-mean-square deviation $\sigma_D$ given
by\footnote{As discussed in
Refs.~\cite{gacmpla,gacgrf98,bignapap},
this form of $\sigma_D$ also implies that
in quantum gravity any measurement that monitors
a distance $D$ for a time $T_{obs}$ is affected by an
uncertainty $\delta D  \ge \sqrt{ {L_{QG} \, c \, T_{obs}}}$.
This must be seen as a minimum uncertainty
that takes only into account the 
quantum and gravitational properties of the measuring
apparatus.
Of course, an even tighter bound can emerge when taking
into account also the quantum and gravitational properties of
the system under observation. According
to the estimates provided in Refs.~\cite{padma,garay}
the contribution to the uncertainty coming from the system
is of the type $\delta D  \ge L_{p}$,
so that the total contribution (summing the system
and the apparatus contributions) might be
of the type $\delta D  \ge L_{p}
+ \sqrt{ {L_{QG} \, c \, T_{obs}}}$.}
\begin{eqnarray}
\sigma_D  \sim \sqrt{ {L_{QG} \, c \, T_{obs}}}
~
\label{gacdlwigner}
\end{eqnarray}
and by
displacement amplitude spectral density $S(f)$ given by
\begin{eqnarray}
S(f) = f^{-1} \sqrt {L_{QG} \, c} ~.
\label{gacspectrwigner}
\end{eqnarray}

Here the scale $L_{QG}$
plays exactly the same role
as in Section~4 (in particular $L_{QG} \equiv {\cal L}_{\beta = 1}$
in the sense of Section~4).
However, seeing $L_{QG}$ as the result of Planck-length
fluctuations occurring at a rate of one per Planck time 
can suggest $L_{QG} \sim L_p$, whereas the different intuition
which has gone into the emergence of $L_{QG}$
in this subsection leaves room for different predictions.
As already emphasized, by mixing elements of quantum mechanics
and elements of gravitation one can only 
conclude that there could 
be some $\sqrt{T_{obs}}$-dependent
irreducible contribution to the uncertainty in the measurement
of distances. One can then guess that space-time foam might
reflect this $\sqrt{T_{obs}}$-dependence and one can parametrize
our ignorance by introducing $L_{QG}$
in the formula $\sqrt{ {L_{QG} \, c \, T_{obs}}}$.
Within such an argument the estimate $L_{QG} \sim L_p$
could only be motivated on dimensional grounds ($L_p$ is
the only length scale available), but
there is no direct estimate of $L_{QG}$ within the argument
advocated in this subsection. We only have
(in the specific sense intended above)
a lower limit on $L_{QG}$ which is set by
the bare analysis using straightforward combination of
elements of ordinary quantum mechanics and elements of
ordinary gravity.
As seen above, this lower limit on $L_{QG}$ 
is set by the minimum allowed value of $1/ M_d$.
Our intuition for $L_{QG}$ might benefit
from trying to establish this minimum allowed value
of $1/ M_d$.
As mentioned, a conservative 
(possibly very conservative)
estimate of this minimum value can be obtained by
enforcing that $M_d$ 
be at least sufficiently small to avoid black hole formation.
In leading order ({\it e.g.}, assuming corresponding
spherical symmetries) this amounts to the requirement
that $M_d < \hbar S_d/(c L_{p}^2)$,
where the length $S_d$ characterizes
the size of the region of space where
the matter distribution associated to $M_d$
is localized.
This observation implies
\begin{eqnarray}
{1 \over M_d}  >
{c L_{p}^2 \over \hbar}
{1 \over S_d} 
~,
\label{estimatea}
\end{eqnarray}
which in turn suggests~\cite{gacmpla}
that $L_{QG} \sim min [L_{p}^2/S_d]$:
\begin{eqnarray}
\delta D  \ge min \sqrt{ {{ {1 \over S_d} 
{L_{p}^2  \, c \, T_{obs} \over 2}}}}
~.
\label{deltagacdlbis}
\end{eqnarray}
Of course, this estimate is very preliminary
since a full quantum gravity would be needed here; in particular,
the way in which black holes were handled
in my argument might have missed important properties
which would become clear only once we have the correct theory.
However, it is nevertheless striking to observe that the
guess $L_{QG} \sim L_{p}$ appears to be very high
with respect to the lower limit on $L_{QG}$
which we are finding from this estimate; in fact,
$L_{QG} \sim L_{p}$ would correspond to the maximum admissible
value of $S_d$ being of order $L_{p}$. 
Since my analysis only holds for devices that can
be treated as approximately rigid\footnote{The fact that
I have included only one contribution
from the quantum properties of the devices, the one associated
with the quantum properties of the motion of the center of mass,
implicitly relies on the assumption that the devices and the bodies
can be treated as approximately rigid. Any non-rigidity of the devices
could introduce additional contributions to the uncertainty
in the measurement of $D$. This is particularly clear
in the case of detector screens and mirrors, whose shape plays
a central role in data analysis. Uncertainties
in the shape (the relative position of different small parts)
of a detector screen or of a mirror
would lead to uncertainties in the measured quantity.
For large devices some level of non-rigidity appears
to be inevitable
in quantum gravity. Causality alone (without any
quantum mechanics) forbids rigid attachment of two bodies
({\it e.g.}, two small parts of a device), but is still consistent
with rigid motion (bodies are not really attached but because
of fine-tuned initial conditions their relative position
and relative orientation are constants of motion).
When Heisenberg's {\it Uncertainty Principle} 
is introduced rigid motion becomes possible only for bodies
of infinite mass (otherwise the relative motion inevitably
has some irreducible uncertainty).
Rigid devices are still available in
ordinary quantum mechanics but they are peculiar devices,
with infinite mass. [Alternatively, in ordinary quantum mechanics
one can take a less fundamental viewpoint on measurement
(which does not appear 
to be natural in the Planck regime~\cite{swok})
in which the trajectory of the different components/parts
of a device are classical because the device is immerged in
a decoherence-inducing environment.]
When both gravitation and quantum mechanics are
introduced rigid devices are no longer available since
the infinite-mass limit is then inconsistent with
the nature of gravitational devices.}
and any non-rigidity could introduce additional
contributions to the uncertainties, it is reasonable to assume
that $max[S_{d}]$ be some small length (small enough that
any non-rigidity would negligibly affect the measurement procedure),
but it appears unlikely that $max[S_{d}] \sim L_{p}$.
This observation might provide some encouragement
for values of $L_{QG}$ smaller than $L_p$,
which after all is the only way to obtain random-walk models
consistent with the data analysis reported
in Refs.~\cite{gacgwi,bignapap}.

Later in this section I will consider a particular class
of estimates for $max[S_{d}]$:
if the correct quantum gravity is such that something
like (\ref{deltagacdlbis}) holds but with $max[S_{d}]$
that depends on $\delta D$ and/or $T_{obs}$, one would have a
different $T_{obs}$-dependence 
(and corresponding $f$-dependence), as I shall
show in one example.

\subsection{Random-walk \space noise \space motivated \space
by \space linear \space deformation \space of \space $~$ 
dispersion relations}

Besides the analysis of the Salecker-Wigner measurement procedure
also the mentioned possibility 
of quantum-gravity-induced deformation of dispersion
relations~\cite{gacgrb,aemn1,gampul,kpoin,lukipapers}
would be consistent with the idea of
random-walk distance fuzziness.
The sense in which this is true is clarified by the
arguments that follow.

Let me start by going back to the general relation
(already discussed in Section~2):
\begin{equation}
c^2{\bf p}^2 \simeq E^2 \left[1 
+ \xi \left({E \over E_{QG}}\right)^{\alpha}
\right] ~.
\label{dispgent}
\end{equation}
Scenarios (\ref{dispgent}) with $\alpha = 1$ are 
consistent with random-walk noise, in the sense that
an experiment involving as
a device (as a probe) a massless particle satisfying the
dispersion relation (\ref{dispgent}) with $\alpha = 1$
would be naturally
affected by a device-induced uncertainty that grows
with $\sqrt{T_{obs}}$. 
From the deformed dispersion relation (\ref{dispgent})
one is led to energy-dependent
velocities~\cite{bignapap}
\begin{equation}
v \simeq c \left[1 
- \left( {1+\alpha \over 2} \right)
 \xi \left({E \over E_{QG}}\right)^{\alpha} \right] ~,
\label{velogen}
\end{equation}
and consequently
when a time $T_{obs}$ has lapsed from the moment in which
the observer
(experimentalist) set off the measurement procedure
the uncertainty in the position of the massless probe
is given by
\begin{equation}
\delta x \simeq  c \, \delta t + \delta v \, T_{obs} 
\simeq c \, \delta t + {1+\alpha \over 2} \, \alpha  \,
{E^{\alpha-1} \, \delta E \over E_{QG}^{\alpha}} c \, T_{obs} ~,
\label{deltagacgen}
\end{equation}
where $\delta t$ is the uncertainty
in the time of emission of the probe, 
$\delta v$ is the uncertainty
in the velocity of the probe,
$\delta E$
is the uncertainty
in the energy of the probe,
and I used the relation
between $\delta v$ and $\delta E$ that follows from (\ref{velogen}).
Since the uncertainty
in the time of emission of a particle and the uncertainty
in its energy are related\footnote{It is well understood
that the $\delta t \, \delta E \ge \hbar$ relation
is valid only in a weaker sense than, say,
Heisenberg's Uncertainty Principle $\delta x \, \delta p \ge \hbar$.
This has its roots in the fact that the time appearing in
quantum-mechanics equations is just a parameter (not an operator),
and in general there is no self-adjoint operator canonically conjugate
to the total energy, if the energy spectrum is bounded 
from below~\cite{pauli,garayclock}.
However, $\delta t \, \delta E \ge \hbar$ 
does relate $\delta t$ intended as uncertainty
in the time of emission of a particle and $\delta E$
intended as uncertainty
in the energy of that same particle.}
by $\delta t \, \delta E \ge \hbar$, Eq.~(\ref{deltagacgen})
can be turned into an absolute bound on 
the uncertainty in the position of the massless
probe when a time $T_{obs}$ has lapsed
from the moment in which
the observer
set off the measurement procedure~\cite{bignapap}
\begin{equation}
\delta x \ge c  {\hbar \over \delta E} 
+ {1+\alpha \over 2} \, \alpha  \,
{E^{\alpha-1} \, \delta E \over E_{QG}^{\alpha}} T_{obs} 
\ge \sqrt{\left({\alpha+\alpha^2 \over 2} \right)
\left( {E \over E_{QG}}\right)^{\alpha-1}
{c^2 \hbar T_{obs} \over E_{QG}}}
~.
\label{deltagacgenfin}
\end{equation}

For $\alpha=1$ the $E$-dependence on the right-hand side of 
Eq.~(\ref{deltagacgenfin}) disappears and one is led again
to a $\delta x$ of the type $(constant) \cdot \sqrt{T_{obs}}$:
\begin{equation}
\delta x \ge \sqrt{{c^2 \hbar T_{obs} \over E_{QG}}}
~.
\label{deltagacgenfinalphaone}
\end{equation}

When massless probes are used in the measurement of a distance $D$
the uncertainty (\ref{deltagacgenfinalphaone}) in the position
of the probe translates directly into an uncertainty on $D$:
\begin{equation}
\delta D \ge \sqrt{{c^2 \hbar T_{obs} \over E_{QG}}}
~.
\label{deltagacgenfinalphaoned}
\end{equation}
This was already observed in Refs.~\cite{aemn1,gacxt,lukipapers}
which considered the implications of deformed dispersion 
relations (\ref{dispgent}) with $\alpha = 1$ 
for the operative definition of distances.

Since deformed dispersion 
relations (\ref{dispgent}) with $\alpha = 1$ 
have led us to the same measurability bound 
already encountered both in the analysis of the Salecker-Wigner 
measurement procedure and the analysis of simple-minded 
random-walk models of quantum space-time fluctuations,
if we assume again that such measurability bounds
emerge in a full quantum gravity as a result of
corresponding quantum fluctuations (fuzziness), we are led
once again to random-walk noise:
\begin{equation}
\sigma_D \sim \sqrt{{c^2 \hbar T_{obs} \over E_{QG}}}
~.
\label{deltagacgenfinalphaonesigmad}
\end{equation}

\subsection{Noise \space motivated \space 
by \space quadratic \space deformation 
\space of \space dispersion \space
relations}

In the preceding subsection I observed that
quantum-gravity deformed dispersion 
relations (\ref{dispgent}) with $\alpha = 1$ 
can also motivate random-walk 
noise $\sigma_D \sim (constant) \cdot \sqrt{T_{obs}}$.
If we use the same line of reasoning that connects a
measurability bound to a scenario for fuzziness
when $\alpha \ne 1$ we appear to
find $\sigma_D \sim {\cal G}(E/E_{QG}) \cdot \sqrt{T_{obs}}$,
where ${\cal G}(E/E_{QG})$ is a ($\alpha$-dependent) function
of $E/E_{QG}$. However, in these cases with $\alpha \ne 1$
clearly the connection between measurability bound
and fuzzy-distance scenario cannot be this simple; in fact,
the energy of the probe $E$ which naturally plays a role in
the context of the derivation of the measurability bound
does not have an obvious counter-part 
in the context of the conjectured fuzzy-distance scenario.

In order to preserve the 
conjectured connection between measurability bounds
and fuzzy-distance scenarios one can be tempted to envision that
if $\alpha \ne 1$ the interferometer noise levels induced
by space-time fuzziness might be of the type~\cite{bignapap}
\begin{equation}
\sigma_D \sim \sqrt{\left({\alpha+\alpha^2 \over 2} \right)
\left( {E^* \over E_{QG}}\right)^{\alpha-1}
{c^2 \hbar T_{obs} \over E_{QG}}} ~,
\label{deltagacgenfinalphamanysigmad}
\end{equation}
where $E^*$ is some energy scale characterizing the physical context
under consideration. [For example, at the intuitive level
one might conjecture that $E^*$
could characterize some sort of energy
density associated with quantum fluctuations of space-time
or an energy scale associated with the masses of the
devices used in the measurement process.]

Since $\alpha \ge 1$ in all quantum-gravity approaches 
believed to support deformed dispersion relations,
it appears likely that
the factor $(E^*/E_{QG})^{\alpha -1}$ would suppress the 
random-walk noise effect in all contexts with $E^* < E_{QG}$.
Besides the case $\alpha = 1$ (linear deformation)
also the case $\alpha = 2$ (quadratic deformation)
deserves special interest since it can emerge quite
naturally in quantum-gravity approaches
(see, {\it e.g.}, Ref.~\cite{thooft}).

\subsection{Noise with $f^{-5/6}$ amplitude spectral density}

In Subsection~8.2 a 
bound on the measurability of distances based on the 
Salecker-Wigner procedure was used as additional motivation
for experimental tests of 
interferometer noise of
random-walk type, with $f^{-1}$ amplitude spectral density
and $\sqrt{T_{obs}}$ root-mean-square deviation.
In this subsection I shall pursue further the observation
that the relevant measurability bound could be derived
by simply insisting that the devices do not turn into black holes.
That observation allowed to derive Eq.~(\ref{deltagacdlbis}),
which expresses the minimum uncertainty $\delta D$
on the measurement of a distance $D$ ({\it i.e.} the measurability
bound for $D$) as proportional to $\sqrt{T_{obs}}$ and 
$\sqrt{1/ S_d}$.
Within that derivation
the minimum uncertainty is obtained 
in correspondence of $max[S_{d}]$, the maximum value of $S_d$
consistent with the structure of the measurement procedure.
I was therefore led to consider how large $S_d$ could be 
while still allowing to disregard any non-rigidity in the quantum
motion of the device (which could introduce additional
contributions to the uncertainties).
Something suggestive of the random-walk
noise scenario emerged
by simply assuming that $max[S_{d}]$ be independent 
of $T_{obs}$ and independent of 
the accuracy $\delta D$ that the observer would wish to achieve.
However, as mentioned, the same
physical intuition that motivates some of the fuzzy space-time
scenarios here considered 
also suggests that quantum gravity might require a novel measurement
theory, possibly involving a new 
type of relation between system and measuring apparatus.
Based on this intuition,
it seems reasonable to contemplate the possibility
that $max[S_{d}]$  might actually depend on $\delta D$.

It is such a scenario that I want to consider in this subsection.
In particular I want to consider the case $max[S_{d}] \sim \delta D$,
which, besides being simple, has the plausible
property that it allows only small devices if the uncertainty
to be achieved is small, while it would allow 
correspondingly larger devices if the observer was
content with a larger uncertainty.
This is also consistent with the idea that elements of non-rigidity
in the quantum motion of extended devices could be neglected
if anyway the measurement is not aiming for great accuracy,
while they might even lead to the most 
significant contributions to the uncertainty if all other sources 
of uncertainty are very small.
[Salecker and Wigner~\cite{wign} would 
also argue that ``large'' devices
are not suitable for very accurate space-time measurements
(they end up being ``in the way'' of the measurement procedure)
while they might be admissible if space-time is being probed 
rather softly.]

In this scenario with $max[S_{d}] \sim \delta D$,
Eq.~(\ref{deltagacdlbis}) takes the form
\begin{eqnarray}
\delta D  \ge \sqrt{ {{ {1 \over S_d}
{L_{p}^2  \, c \, T_{obs} \over 2}}}}
\ge \sqrt{ {{ L_{p}^2 \, c \, T_{obs} \over 2\,\, \delta D }}}
~,
\label{predeltagactwothird}
\end{eqnarray}
which actually gives
\begin{eqnarray}
\delta D  \ge \left( {{1 \over 2} L_{p}^2 \, 
c \, T_{obs} }\right)^{1/3}
~.
\label{deltagactwothird}
\end{eqnarray}
As done with 
the other measurability bounds,
I have proposed~\cite{gacgwi,bignapap}
to take Eq.~(\ref{deltagactwothird}) as motivation
for the investigation of 
a corresponding fuzziness scenario characterised by
\begin{eqnarray}
\sigma_D  \sim \left( {\tilde L}_{QG}^2 \, 
c \, T_{obs} \right)^{1/3}
~.
\label{sigmagactwothird} 
\end{eqnarray}
Notice that in this equation I replaced $L_{p}$
with a generic length scale ${\tilde L}_{QG}$, 
since it is possible that the heuristic
argument leading to Eq.~(\ref{sigmagactwothird}) 
might have captured
the qualitative structure of the phenomenon while providing 
an incorrect estimate of the relevant length scale.
Also notice that Eq.~(\ref{deltagactwothird})
has the same form as the relations emerged in
other measurability analyses~\cite{ng,karo},
even though those analyses adopted a
very different viewpoint
(and even the physical interpretation of the elements
of Eq.~(\ref{deltagactwothird}) was different,
as explained in the next section).

As observed in Refs.~\cite{gacgwi,bignapap}
the $T_{obs}^{1/3}$ dependence of $\sigma_D$ is associated with
displacement amplitude spectral density with $f^{-5/6}$ behaviour:
\begin{eqnarray}
{\cal S}(f) = f^{-5/6} ({\tilde L}_{QG}^2 \, c)^{1/3}
~.
\label{no3nspectr}
\end{eqnarray}
Therefore the measurability analyses discussed in this subsection
provides motivation for the investigation of the case $\beta=5/6$
(using again the notation set up
in Section~4).

\section{ABSOLUTE MEASURABILITY BOUND FOR $~~~~~$
THE AMPLITUDE OF A GRAVITY WAVE}

The bulk of this Article 
(presented in the previous three sections) 
concerns the implications
of distance fuzziness for interferometry.
Various scenarios for distance fuzziness were motivated either
by a general Wheeler-Hawking-inspired phenomenological
parametrization or by intuitive arguments based on
the possibility of quantum-gravity-induced deformations
of dispersion relations or quantum-gravity\footnote{My
observations within the
Salecker-Wigner setup do pertain to the quantum-gravity
realm because I took into account the gravitational
properties of the devices and I also, like Salecker and Wigner,
removed the assumption of classicality of the devices.
If one was only putting together some properties
of gravitation and quantum mechanics one could at best probe
a simple limiting behaviour of quantum gravity, but by removing
one of the conceptual ingredients of
ordinary quantum mechanics it is 
plausible that we get a glimpse of a true property of quantum
gravity. The Salecker-Wigner study~\cite{wign} 
(just like the Bohr-Rosenfeld analysis~\cite{rose})
suggests that among the conceptual elements of
quantum mechanics the one that is most likely
(although there are of course no guarantees)
to succumb to the unification of gravitation and quantum
mechanics is the requirement for devices to be treated
as classical.}
distance-measurability analyses
within the Salecker-Wigner setup.
My observation that distance fuzziness would be
felt by interferometers as a fundamental
additional source of noise
({\it i.e.} as some sort of fundamental
source of stochastic gravity-wave background)
also implies that, if indeed quantum gravity
hosts distance fuzziness,
there would be a quantum-gravity induced
bound on the measurability
of gravity waves.
This section is 
parenthetical, within the logical line of this Article,
in the sense that
I will assume in this section that there
is no distance fuzziness. The objective is one of
showing that even without distance fuzziness it appears
that the measurability
of gravity waves should be limited in quantum gravity.

The strategy I will use to derive this bound is an
adaptation of the Salecker-Wigner framework to the analysis
of gravity-wave measurability.
Basically, while the Salecker-Wigner framework 
concerns the measurement of a distance $D$, I shall here
apply the same reasoning to the measurement
of ``distance displacements'' in interferometers arms
(of length $L$)
of the type that could be induced by a gravity wave.

Having clarified in which sense this section represents a deviation
from the main bulk of observations reported in the present Article,
let me start the discussion
by reminding the reader of the fact that, as 
already mentioned in Section~2,
the interference pattern generated by a modern interferometer
can be remarkably sensitive to changes in the positions of the mirrors
relative to the beam splitter, and is therefore
sensitive to gravitational waves (which, as described in
the {\it proper reference frame}~\cite{saulson},
have the effect of changing these relative positions).
With just a few lines of simple algebra
one can show that an ideal gravitational wave
of amplitude $h$ and reduced\footnote{I report
these results
in terms of reduced wavelengths $\lambda^{o}$
(which are related to the wavelengths $\lambda$ by 
$\lambda^{o}= \lambda/(2 \pi)$) in order to avoid 
cumbersome factors of $\pi$ in some of the formulas.}
wavelength $\lambda^{o}_{gw}$
propagating along the direction orthogonal to
the plane of the interferometer would cause a change in the
interference pattern as for a phase
shift of magnitude $\Delta \phi = D_L/ \lambda^{o}$,
where $\lambda^{o}$ is the reduced
wavelength of the laser beam used in
the measurement procedure and~\cite{saulson,qigwdbook}
\begin{eqnarray}
D_L \sim 2 \, h \,  \lambda^{o}_{gw} \,
\left| \sin \left( {L \over
2 \lambda^{o}_{gw}} \right) \right|
~,
\label{dleq}
\end{eqnarray}
is the
magnitude of the change caused by the gravitational wave in the length 
of the arms of the interferometer. 
(The changes in the lengths of the two arms
have opposite sign~\cite{saulson}.)
 
As already mentioned in Section~2,
modern techniques allow to construct
gravity-wave interferometers with truly remarkable sensitivity;
in particular, at least for
gravitational waves with $\lambda^{o}_{gw}$ of
order $10^3 Km$, the next LIGO/VIRGO generation of detectors
should be sensitive to $h$ as low as $3 \cdot 10^{-22}$.
Since $h \sim 3 \cdot 10^{-22}$ causes a $D_L$ of order $10^{-18}m$
in arms lengths $L$ of order $3 Km$, it is not surprising 
that in the analysis of gravity-wave interferometers,
in spite of their huge size,
one ends up having to take into account~\cite{saulson}
the type of quantum effects usually significant only for the study of
processes at or below the atomic scale.
In particular, there is the so-called {\it standard quantum limit}
on the measurability of $h$ that results 
from the combined minimization
of {\it photon shot noise} and {\it radiation pressure noise}.
While a careful discussion of these two noise sources 
(which the interested
reader can find in Ref.~\cite{saulson})
is quite insightful, here I shall rederive
this {\it standard quantum limit} in an alternative\footnote{While
the {\it standard quantum limit} can be equivalently obtained
either from the combined minimization
of {\it photon shot noise} and {\it radiation pressure noise}
or from the application of Heisenberg's uncertainty
principle to the position and momentum of the mirror,
it is this author's opinion that there might
actually be a fundamental
difference between the two derivations.
In fact, it appears (see, {\it e.g.}, Ref.~\cite{jare1}
and references therein)
that the limit obtained through combined minimization
of {\it photon shot noise} and {\it radiation pressure noise}
can be violated by careful exploitation of the properties
of squeezed light, whereas the 
limit obtained through the application of Heisenberg's uncertainty
principle to the position and momentum of the mirror
is truly fundamental.}
and straightforward manner (also discussed in Ref.~\cite{qigwdbook}),
which relies on the application of Heisenberg's uncertainty
principle to the position and momentum of a mirror relative
to the position of the beam splitter.
This can be done along the lines of my analysis
of the
Salecker-Wigner procedure for the measurement of distances.
Since the mirrors and the beam splitter 
are macroscopic, and therefore
the corresponding momenta and velocities are related 
non-relativistically, Heisenberg's uncertainty principle
implies that~\footnote{Note that in the setup 
of gravity-wave interferometers the test masses
are required to be free-falling~\cite{saulson}.
In such a context the type
of observations reported in Ref.~\cite{anos} is not only
inadequate for in-principle analyses of measurability
in the full quantum-gravity regime but in most cases,
as a result of the free-fall requirement, it will
also be inapplicable in the ordinary context
of present-day interferometers.}
\begin{eqnarray}
\delta x \, \delta v \ge {\hbar \over 2} \left({1 \over M_m} 
+ {1 \over M_b} \right) \ge {\hbar \over 2 M_m}
~,
\label{deltaheis}
\end{eqnarray}
where $\delta x$ and $\delta v$ are the uncertainties in the
relative position and relative velocity,
$M_m$ is the mass of the mirror, $M_b$ is the mass
of the beam splitter. [Again, the relative motion is
characterised by the {\it reduced mass}, which is
given in this case by $(1/M_m+1/M_b)^{-1}$.] 

Clearly, the high precision of the planned measurements
requires~\cite{saulson,qigwdbook} that the
position of the mirrors
be kept under control during the whole time $2L/c$
that the beam spends in between the arms of
the detector before superposition.
When combined with (\ref{deltaheis}) this leads to the finding
that, for any given value of $M_m$,
the $D_L$ induced by the gravitational wave can be measured
only up to an irreducible uncertainty,
the so-called {\it standard quantum limit}:
\begin{eqnarray}
\delta D_L \ge  \delta x + \delta v \, 2 {L \over c}
\ge \delta x + {\hbar L \over c M_m \delta x} 
\ge  \sqrt{{\hbar L \over c M_m }}
~.
\label{deltawign}
\end{eqnarray}

The case of gravity-wave measurements is a canonical
example of my general argument that the infinite-mass 
classical-device limit underlying ordinary quantum mechanics
is inconsistent with the nature of gravitational measurements.
As the devices get more and more massive they not only
increasingly disturb the gravitational/geometrical observables, 
but eventually 
(well before reaching the infinite-mass limit)
they also render impossible~\cite{gacmpla,gacgrf98} 
the completion of the procedure of measurement 
of gravitational observables.
In trying to asses how this observation affects
the measurability of the properties of a gravity wave
let me start by combining Eqs.~(\ref{dleq})
and (\ref{deltawign}):
\begin{eqnarray}
\delta h = \delta \left({D_L \over L}\right)
= h {\delta D_L \over D_L}
\ge {\sqrt{{\hbar L \over c M_m }} \over
2 \,  \lambda^{o}_{gw} \,
\left| \sin \left( {L \over 2 \lambda^{o}_{gw}} 
\right) \right| }
~.
\label{deltawignh}
\end{eqnarray}
In complete analogy with some of the observations
made in Section~3 concerning the measurability of distances,
I observe that, when  
gravitational effects are taken into account,
the limit of infinite mirror mass is of course inadmissable.
At the very least $M_m$
must be small enough that
the mirror does not turn into a black hole.\footnote{This is of 
course a very conservative bound, since
a mirror stops being useful as a device well before it turns into
a black hole, but even this conservative approach leads 
to an interesting conclusion.}
In order for the mirror not to be a black hole
one requires $M_m < \hbar S_m/(c L_{p}^2)$,
where $S_m$ is the {\it size} of 
the region of space occupied by the mirror.
This observation combined with (\ref{deltawignh})
implies that one would have obtained a bound on
the measurability of $h$
if one found a maximum allowed mirror size $S_m$.
In estimating this maximum $S_m$
one can be easily led to some
extreme and incorrect assumptions.
In particular, one could suppose
that in order to achieve a sensitivity to $D_L$ as
low as $10^{-18} m$ it might be necessary
to ``accurately position''
each $10^{-36} m^2$ surface element of the mirror.
If this was
really necessary,
our line of argument would then lead to a rather large
measurability bound.
Fortunately, the phase of the wavefront of the reflected
light beam is determined
by the average position of all the atoms across the
beam's width,
and microscopic irregularities in the structure of
the mirror only lead to scattering of a small fraction
of light out of the beam.
This suggests that in our analysis the size of the mirror
should be assumed to be of the order of the width of the 
beam~\cite{saulson}.
So $S_m$ cannot be too small, but on the other hand
in light of this observation, and taking into account
the in-principle nature\footnote{For the
gravitational waves to which LIGO/VIRGO will be most sensitive,
which have $\lambda^{o}_{gw}$ of
order $10^3 Km$, the
requirement $S_m < \lambda^{o}_{gw}$
simply states that the size of mirrors
should be smaller than $10^3 Km$.
This bound might appear very conservative, but I am
trying to establish an in-principle
limitation on the measurability of $h$.
Since such a bound was not previously established,
in this first study I just want to clarify that the
bound exists, rather than dwell on the exact
magnitude of the bound. I therefore prefer
to be very conservative in my estimates.}
of the analysis I am performing,
it is clear that $S_m$
could not be too large either, and in particular
it appears safe to assume that $S_m$
should be smaller than the $\lambda^{o}_{gw}$
of the gravity wave
which one is planning to observe.
If $S_m$ is indeed the width of the beam 
(and therefore the effective size of the mirror),
then one must exclude the possibility $S_m > \lambda^{o}_{gw}$
because otherwise the
same gravity wave which
one is intending to observe would cause phenomena preventing
the proper completion of the measurement
procedure ({\it e.g.} deforming the mirror and leading to a nonlinear
relation between $D_L$ and $h$).
The conservative bound $S_m < \lambda^{o}_{gw}$
also appears o be safe with respect to the
expectations of another type of intuition,
usually resulting from experience
with table-top interferometers.
Within this assumption
one is always tempted to think of the mirror as {\it attached}
to a very massive body.
Even setting aside the limitations on this type of idealized
attachements that are set by the uncertainty principle and
causality, it appears that the bound $S_m < \lambda^{o}_{gw}$
should be safe because of the requirement that
the mirror be free-falling. [It actually
seems extremely conservative to just demand of such
a free-falling interferometer
mirror that the sum of its mass and the mass of
any body ``attached'' to it should not exceed the mass
of a black hole of size $\lambda^{o}_{gw}$.]

In summary, it looks very
safe to assume that $M_m$ should be smaller 
than $\hbar \lambda^{o}_{gw} /(c L_{p}^2)$, and
this can be combined with (\ref{deltawignh})
to obtain the measurability bound 
\begin{eqnarray}
\delta h > {L_{p} \over 2 \, \lambda^{o}_{gw}} 
{\sqrt{{ L/ \lambda^{o}_{gw} }} \over
\left| \sin \left( {L \over 2 \lambda^{o}_{gw}} 
\right) \right| }
~.
\label{hbound}
\end{eqnarray}
This result not only sets a
lower bound on the measurability of $h$
with given arm's length $L$, but also
encodes an absolute ({\it i.e.}
irrespective of the value of $L$)
lower bound, as a result of the fact that
the function $\sqrt{x}/|\sin (x/2)|$
has an absolute minimum: $min[\sqrt{x}/\sin (x/2)] \sim 1.66$.
This novel measurability bound
is a significant departure from the principles of ordinary
quantum mechanics, especially in light of the fact that
it describes a limitation on the measurability of a single
observable (the amplitude $h$ of a gravity wave),
and that this limitation turns out to depend on the value (not the 
associated uncertainty) of another observable
(the reduced wavelength $\lambda^{o}_{gw}$
of the same gravity wave).
It is also significant that this new bound (\ref{hbound})
encodes an aspect of a novel type of interplay
between system and measuring apparatus in quantum-gravity regimes;
in fact, in deriving (\ref{hbound}) a crucial role was played by the
fact that in accurate measurements of gravitational/geometrical 
observables it is no longer possible~\cite{gacgrf98} 
to advocate an idealized description of the devices. 

Also the $T_{obs}$-dependent 
bound on the measurability of distances which I reviewed
in Section~3 encodes a
departure from ordinary quantum mechanics and 
a novel type of interplay
between system and measuring apparatus,
but the bound (\ref{hbound})
on the measurability of the amplitude of 
a gravity wave (which is one of the new results reported
in the present Article) should provide even stronger motivation
for the search of formalisms in which quantum gravity 
is based on a new mechanics,
not exactly given by ordinary quantum mechanics.
In fact, while one might still hope
to find alternatives to the Salecker-Wigner measurement
procedure that allow to measure distances evading the 
corresponding measurability bounds,
it appears hard to imagine that there could be 
anything (even among ``gedanken laboratories'')
better than an interferometer for measurements
of the amplitude of a gravity wave.

It is also important to realize that the bound (\ref{hbound})
cannot be obtained by just assuming
that the Planck length $L_{p}$
provides the minimum uncertainty for distances (and distance
variations).
In fact, if the only limitation was $\delta D_L \ge L_{p}$
the resulting
uncertainty on $h$, which I denote with $\delta h^{(L_{p})}$,
would have the property
\begin{eqnarray}
min [\delta h^{(L_{p})}] = min \left[ {L_{p} \over 2 
\, \lambda^{o}_{gw}
\left| \sin \left( {L \over 2 \lambda^{o}_{gw}} 
\right) \right|} \right] = {L_{p} \over 2 
\, \lambda^{o}_{gw}}
~,
\label{hboundlp}
\end{eqnarray}
whereas, exploiting the above-mentioned 
properties of the function $\sqrt{x}/|\sin (x/2)|$,
from (\ref{hbound}) one finds\footnote{I am here
(for ``pedagogical'' purposes) somewhat simplifying the comparison
between $\delta h$ and $\delta h^{(L_{p})}$.
As mentioned, in principle one should take into account
both uncertainties inherent in the ``system'' under observation,
which are likely to be characterized exclusively by
the Planck-length bound, and uncertainties coming from 
the ``measuring apparatus'', which might easily involve
other length (or time) scales besides the Planck length.
It would therefore be proper to compare $\delta h^{(L_{p})}$,
which would be the only contribution present in the conventional
idealization of ``classical devices'', with the 
sum $\delta h + \delta h^{(L_{p})}$,
which, as appropriate for quantum gravity,
provides a sum of system-inherent uncertainties plus
apparatus-induced uncertainties.}
\begin{eqnarray}
min [\delta h]  > min \left[
{L_{p} \over 2 \, \lambda^{o}_{gw}} 
{\sqrt{{ L/ \lambda^{o}_{gw} }} \over
\left| \sin \left( {L \over 2 \lambda^{o}_{gw}} 
\right) \right| } \right]
> min [\delta h^{(L_{p})}] 
~.
\label{minhbound}
\end{eqnarray}
In general,  the dependence of $\delta h^{(L_{p})}$
on $\lambda^{o}_{gw}$
is different from the one of $\delta h$.
Actually, in light of the comparison of
(\ref{hboundlp}) with (\ref{minhbound})
it is amusing to observe that
the bound (\ref{hbound}) could be seen as
the result of a minimum length $L_{p}$ combined
with an $\lambda^{o}_{gw}$-dependent correction.
This would be consistent with some of the ideas mentioned
in Section~3
(the energy-dependent
effect of {\it in vacuo} dispersion and the corresponding
proposal (\ref{deltagacgenfinalphaonesigmad}) for distance
fuzziness)
in which the magnitude of the quantum-gravity effect
depends rather sensitively on some energy-related 
aspect of the problem under investigation
(just like $\lambda^{o}_{gw}$ for the
gravity wave).

It is easy to verify that
the bound (\ref{hbound}),
would not observably affect the operation of even the most 
sophisticated planned interferometers.
However, in the spirit of what I did in the previous sections
considering the operative definition of distances, also
for the amplitudes of gravity waves the fact that we have
encountered
an obstruction in the measurement analysis based on ordinary
quantum mechanics
(and the fact that by mixing gravitation and quantum
mechanics we have obtained some intuition for novel qualitative
features of such gravity-wave amplitudes in quantum gravity)
could be used as starting point for
the proposal of novel quantum-gravity effects
possibly larger than the estimate (\ref{hbound}).
Although possibly very interesting,
these fully quantum-gravity scenarios 
for the properties of gravity-wave amplitudes
will not be explored in these notes.
I just want to observe that the strain sensitivity 
($S_h(f) \equiv S(f)/L$) of 
order $10^{-22}/\sqrt{H\!z}$ 
which is soon going to be achieved by 
several detectors~\cite{ligo,virgo,nautilus,stringcoreviews}
corresponds to a rather natural scale for
a fundamental quantum-gravity-induced 
stochastic-gravity-wave-like noise;
in fact, $10^{-22}/\sqrt{H\!z} \simeq \sqrt{L_p/c}$.

\section{RELATIONS WITH OTHER \hfil $~$ \hfil $~$ \hfil $~$
QUANTUM-GRAVITY APPROACHES}

In this section I comment on the connections
and the differences between some of the ideas
that I reviewed in these notes and
other quantum-gravity ideas.

\subsection{Canonical Quantum Gravity}

One of the most popular quantum-gravity approaches
is the one in which the ordinary canonical formalism
of quantum mechanics is applied to (some formulation of)
Einstein's Gravity.
In spite of the fact that~\cite{gacgrf98}
some of the
observations reviewed in the previous sections
suggest that quantum gravity should require a new mechanics,
not exactly given by ordinary quantum mechanics,
it is very interesting\footnote{I am here taking a viewpoint
that might be summarized rephrasing a comment by B.S.~De~Witt
in Ref.~\cite{dewitt}. While some of the arguments reviewed here
appear to indicate that ordinary quantum mechanics cannot suffice for
quantum gravity, it is still plausible that the language of ordinary
quantum mechanics might be a useful tool for the description
of its own demise.
This would be analogous to something we have learned in the study
of special relativity: one could~\cite{dewitt} insist
on describing the observed Lorentz-Fitzgerald contraction as the
result of relativistic modifications in the force law between atoms,
but in order to capture the true essence of the new regime
it is necessary to embrance the new conceptual framework
of special relativity.}
that some of the phenomena
considered in the previous sections
have also emerged in 
studies of canonical quantum gravity.

As mentioned, the most direct connection was found in
the study reported in Ref.~\cite{gampul},
which was motivated by Ref.~\cite{gacgrb}.
In fact, Ref.~\cite{gampul} shows that the popular
canonical/loop quantum gravity~\cite{canoloop}
admits the phenomenon of deformed dispersion
relations, with the deformation going linearly with
the Planck length.

Concerning the bounds on the measurability of distances
it is probably fair to say that the situation
in canonical/loop quantum gravity is not yet clear
because the present formulations
do not appear to lead to
a compelling candidate ``length operator.''
This author would like to interpret
the problems associated
with the length operator as an indication that 
perhaps something unexpected
might actually emerge in canonical/loop quantum gravity
as a length operator,
possibly something with properties fitting the intuition
of some of the scenarios for fuzzy distances
which I reviewed.
Actually, the random-walk 
space-time fuzziness model
might have a (somewhat weak, but intriguing) 
connection with ``quantum mechanics applied to gravity''
at least to the level seen by 
comparison with the scenario discussed 
in Ref.~\cite{fotinilee}, which was motivated by 
the intuition that is emerging from investigations of 
canonical/loop quantum gravity.
The ``moves'' of Ref.~\cite{fotinilee} share many of the
properties of the ``random steps'' of the random-walk models
here considered. 

\subsection{Critical and non-critical String Theories}

Unfortunately, in the popular quantum-gravity approach
based on critical superstrings\footnote{As already mentioned
the mechanics of critical superstrings is just an ordinary
quantum mechanics.
All of the new structures
emerging in this exciting formalism are the result of
applying ordinary quantum mechanics to the dynamics of
extended fundamental objects,
rather than point-like objects (particles).} 
not many results have been derived 
concerning directly the quantum properties of space-time.
Perhaps the most noticeable such results are the ones on
limitations on the measurability of distances emerged in the
scattering analyses reported in Refs.~\cite{venekonmen,dbrscatt},
which I already mentioned.

A rather different picture is emerging (through the difficult
technical aspects
of this rich formalism) in {\it Liouville} (non-critical)
strings~\cite{emn}, whose development was partly motivated 
by intuition concerning the quantum-gravity vacuum
that is rather close to the one traditionally associated
with the mentioned works of
Wheeler and Hawking.
Evidence has been found~\cite{aemn1} in {\it Liouville} 
strings supporting the validity  
of deformed dispersion relations, with the 
deformation going linearly with the Planck/string length.
In the sense clarified in Subsection~8.3 this approach might
also host a bound on the measurability of distances which grows
with $\sqrt{T_{obs}}$.

\subsection{Other types of measurement analyses}

Because of the lack of experimental input,
it is not surprising that
many authors have been seeking some
intuition on quantum gravity
by formal analyses of the ways in which
the interplay between gravitation and quantum mechanics
could affect measurement procedures.
A large portion of these analyses
produced a ``$min [\delta D]$'' with $D$ denoting
a distance; however, the same type of notation
was used for structures defined in significantly
different ways.
Also different meanings have been given by different
authors to the statement ``absolute bound on the measurability
of an observable.''
Quite important for the topics here discussed
are the differences (which might not be
totally transparent as a result of this unfortunate
choice of overlapping notations)
between the approach advocated 
in Refs.~\cite{gacgwi,gacmpla,gacgrf98,bignapap}
and the approaches advocated in Refs.~\cite{wign,diosi,ng,karo}.
In Refs.~\cite{gacgwi,gacmpla,gacgrf98,bignapap} ``$min [\delta D]$''
denotes an absolute limitation on the measurability of a distance $D$.
The studies~\cite{wign,diosi,karo} 
analyzed the interplay of gravity and quantum mechanics
in defining a net of time-like geodesics, and 
in those studies ``$min [\delta D]$''
characterizes the maximum ``tightness'' achievable
for the net of time-like geodesics.
Moreover, 
in Refs.~\cite{wign,diosi,ng,karo} it was required that
the measurement procedure should not affect/modify the
geometric observable being measured, 
and ``absolute bounds on the measurability''
were obtained in this specific sense.
Instead, in Refs.~\cite{gacmpla,gacgrf98,bignapap}
it was envisioned that
the observable which is being measured might depend
also on the devices (the underlying view is that observables
in quantum gravity would always be, in a sense, shared properties
of ``system'' and ``apparatus''),
and it was only required that
the nature of the devices be consistent with the various
stages of the measurement procedure (for example if a device
turned into a black hole some of the exchanges
of signals needed for the measurement
would be impossible).
The measurability bounds
of Refs.~\cite{gacmpla,gacgrf98,bignapap}
are therefore to be intended from this
more fundamental perspective, and this is crucial for the
possibility that these measurability bounds be associated
to a fundamental quantum-gravity mechanism for ``fuzziness''
(quantum fluctuations of space-time).
The analyses reported in Refs.~\cite{wign,diosi,ng,karo} 
did not include any reference to fuzzy space-times
of the type operatively defined in terms of stochastic processes
in Section~4 (and in Ref.~\cite{bignapap}).

The more fundamental nature of the bounds obtained
in Refs.~\cite{gacmpla,gacgrf98,bignapap}
is also crucial for the arguments suggesting
that quantum gravity might require a new mechanics,
not exactly given by ordinary quantum mechanics.
The analyses reported
in Refs.~\cite{wign,diosi,ng,karo} 
did not include any reference to this possibility.

Having clarified that there is 
a ``double difference'' (different ``$min$'' and 
different ``$\delta D$'')
between the meaning of $min [\delta D]$ 
adopted in Refs.~\cite{gacgwi,gacmpla,gacgrf98,bignapap}
and the meaning of $min [\delta D]$ adopted
in Refs.~\cite{wign,diosi,ng,karo}, it is however important to 
notice that the studies reported
in Refs.~\cite{diosi,ng,karo} 
were among the first studies which showed how in some aspects
of measurement analysis the Planck length might appear
together with other length scales in the problem.
For example, a quantum-gravity effect 
naturally involving something
of length-squared dimensions might not necessarily 
go like $L_{p}^2$: in some cases it could go
like $\Lambda L_{p}$, with $\Lambda$ some other 
length scale in the problem.

Interestingly, the
analysis of the interplay of gravity and quantum mechanics
in defining a net of time-like geodesics reported
in Ref.~\cite{diosi}
concluded that the maximum ``tightness'' achievable 
for the geodesics would be characterized
by $\sqrt{L_{p}^2 R^{-1} s}$,
where $R$ is the radius of the (spherically symmetric) clocks
whose world lines define the network of geodesics,
and $s$ is the characteristic distance scale 
over which one is intending to define such a network.
The $\sqrt{L_{p}^2 R^{-1} s}$ maximum tightness discussed 
in Ref.~\cite{diosi} is formally
analogous to Eq.~(\ref{deltagacdlbis}),
but, as clarified above, this ``maximum tightness''
was defined in a very different (``doubly different'') way,
and therefore the two proposals have 
completely different physical implications.
Actually, in Ref.~\cite{diosi} it was also stated that 
for a single geodesic distance (which might be closer
to the type of distance measurability analysis reported
in Refs.~\cite{gacmpla,gacgrf98,bignapap})
one could achieve accuracy significantly better
than the formula $\sqrt{L_{p}^2 R^{-1} s}$,
which was interpreted in Ref.~\cite{diosi} as a direct
result of the structure of a network of geodesics.

Relations of the type $min [\delta D] \sim (L_{p}^2 D)^{(1/3)}$,
which are formally analogous to Eq.~(\ref{deltagactwothird}),
were encountered in the analysis of 
maximum tightness achievable for 
a geodesics network reported in Ref.~\cite{karo} 
and in the analysis of measurability of distances 
reported in Ref.~\cite{ng}.
Although once again the definitions of ``$min$''
and ``$\delta D$'' used in these studies
are completely different from the ones relevant for 
the ``$min [\delta D]$'' of Eq.~(\ref{deltagactwothird}).

\section{QUANTUM GRAVITY, NO STRINGS (OR LOOPS) ATTACHED}

Some of the arguments reviewed in these notes appear
to suggest that quantum gravity might require a mechanics
not exactly of the type of ordinary quantum mechanics.
In particular, the new mechanics might have to
accommodate a somewhat different relationship 
(in a sense, ``more democratic'')
between ``system''
and ``measuring apparatus'', and should take into account
the fact that the limit in which the apparatus behaves classically
is not accessible once gravitation is turned on.
The fact that the most popular
quantum-gravity approaches, including critical
superstrings
and canonical/loop quantum gravity,
are based on ordinary quantum mechanics but
seem inconsistent with a
correspondence between formalism and measurability bounds
of the type sought and found in non-gravitational
quantum mechanics (through the work of Bohr, Rosenfeld, Landau,
Peierls, Einstein, Salecker, Wigner and many others),
represents, in this author's humble opinion,
one of the outstanding problems of these approaches.
Still, it is of great importance for quantum-gravity research
that these approaches continue to be pursued very aggressively:
they might eventually encounter along their development
unforeseeable answers to these questions or else,
as they are ``pushed to the limit'',
they might turn out to fail in a way that provides insight on the
correct theory.
However, the observations pointing us
toward deviations from ordinary
quantum-mechanics could provide
motivation for the parallel development
of alternative quantum-gravity approaches.
But how could we envision quantum gravity with no strings
(or``canonical loops'') attached?
More properly, how can we devise a new mechanics when we have no
direct experimental data on its structure?
Classical mechanics was abandoned for quantum mechanics only after
a relatively long period of analysis of 
physical problems such as black-body spectrum 
and photoelectric effect which contained very relevant information.
We don't seem to have any such insightful physical problem.
At best we might have identified the type of conceptual issues
which Mach had discussed with respect to Newtonian physics.
It is amusing to
notice that the analogy with Machian conceptual analyses might
actually be quite proper, since at the beginning of this century
we were forced to renounce to the comfort of the reference
to ``absolute space'' and now that we are reaching the end of this
century we might be forced to renounce to the comfort of
an idealized classical measuring apparatus.

Our task is that much harder in light of the fact that 
(unless something like large extra dimensions is verified in Nature) 
we must make a gigantic leap from the energy scales we presently
understand to Planckian energy scales.
While of course it is not impossible that we eventually do come
up with the correct recipe for this gigantic jump,
one less optimistic strategy that might be worth pursuing is the one
of trying to come up with some effective theory 
useful for the description of
new space-time-related phenomena occurring in an energy-scale range
extending from somewhere not much above presently achievable 
energies up to somewhere safely below the Planck scale.
These theories might provide guidance to experimentalists,
and in turn (if confirmed by experiments) might provide a useful 
intermediate step toward the Planck scale.
For those who are not certain that we can make a lucky guess of
the whole giant step toward the Planck
scale\footnote{Understandably,
some are rendered prudent by the realization that
the ratio between the Planck scale and 
the energy scales we are probing with
modern particle colliders is so big that
it is, for example, comparable (within a couple of orders of magnitude)
to the ratio between the average Earth-Moon distance 
%The average distance between the Earth and the Moon is 382176 km. 
%Bohr radius \sim 5*10{-11}m
and the Bohr radius.}
this strategy might provide a possibility to eventually get to
the Planck regime only after a (long and painful) series
of intermediate steps.
Some of the ideas discussed in the previous sections can be seen
as examples of this strategy. In this section I collect
additional relevant material.

\subsection{A low-energy effective theory
of quantum gravity}

While the primary emphasis has been on experimental tests of
quantum-gravity-motivated candidate phenomena,
some of the arguments 
(which are based on Refs.~\cite{gacmpla,gacgrf98,bignapap})
reviewed in these lecture notes
can be seen as attempts to identify 
properties that one could demand of a theory
suitable for a first stage of partial
unification of gravitation and quantum mechanics.
This first stage of partial unification
would be a low-energy effective theory capturing
only some rough features of quantum gravity.
In particular, as discussed in
Refs.~\cite{gacxt,gacgrf98,bignapap},
it is plausible that 
the most significant implications
of quantum gravity for low-energy (large-distance)
physics might be associated with
the structure of the non-trivial ``quantum-gravity vacuum''.
A satisfactory picture of this vacuum
is not available at present, and therefore
we must generically characterize 
it as the appropriate new concept that in quantum gravity
takes the place of the ordinary concept of ``empty space'';
however, it is plausible that some of 
the arguments by Wheeler, Hawking
and followers,
attempting to develop an intuitive description
of the quantum-gravity vacuum, might have captured 
at least some of its actual properties.
Therefore the experimental investigations of space-time foam
discussed in some of the preceding sections could
be quite relevant
for the search of a theory
describing a first stage of partial
unification of gravitation and quantum mechanics.

Other possible elements for the search of such a theory
come from studies suggesting that this unification might
require a new (non-classical) concept
of measuring apparatus and 
a new relationship between measuring apparatus
and system. 
I have reviewed some of the relevant
arguments~\cite{gacmpla,gacgrf98}
through the discussion of the
Salecker-Wigner setup for the measurement of distances,
which manifested the problems associated with
the infinite-mass classical-device limit.
As mentioned,
a similar conclusion was already drawn in the context
of attempts (see, {\it e.g.}, Ref.~\cite{bergstac})
to generalize to
the study of the measurability of gravitational fields
the famous Bohr-Rosenfeld analysis~\cite{rose}
of the measurability of the electromagnetic field.
It seems reasonable to explore the possibility that
already the first stage of partial
unification of gravitation and quantum mechanics
might require a new mechanics.
A (related) plausible feature
of the correct effective low-energy theory
of quantum-gravity is (some form of)
a novel bound on the measurability of distances.
This appears to be an inevitable consequence of
relinquishing the idealized methods of
measurement analysis that rely
on the artifacts of the infinite-mass
classical-device limit.
If indeed one of these novel measurability
bounds holds in the physical world, and if indeed the
structure of the quantum-gravity vacuum
is non-trivial and involves space-time fuzziness,
it appears also plausible that this two features be related,
{\it i.e.} that the fuzziness of space-time would be
ultimately responsible for the measurability bounds.
It is also plausible~\cite{gacxt,gacgrf98}
that an effective large-distance description of some
aspects of quantum gravity might involve quantum symmetries
and noncommutative geometry (while at the
Planck scale even more novel geometric structures might be required).

The intuition emerging from these considerations on
a novel relationship between measuring apparatus
and system and by a Wheeler-Hawking picture
of the quantum-gravity vacuum
has not yet been implemented in a fully-developed new
formalism describing the first stage of partial
unification of gravitation and quantum mechanics,
but one can use this emerging intuition for rough estimates
of certain candidate quantum-gravity effects.
Some of the theoretical estimates that I reviewed in the
preceding sections, particularly the ones on 
distance fuzziness, can be seen as examples of this.

Besides the possibility of direct experimental tests
(such as some of the ones here reviewed),
studies of low-energy effective quantum-gravity models 
might provide a perspective on quantum gravity
that is complementary with respect to the
one emerging from approaches based on proposals
for a one-step full unification of gravitation and
quantum mechanics.
On one side of this complementarity there are the
attempts to find a low-energy effective quantum gravity
which are necessarily driven by intuition based on
direct extrapolation from known physical regimes;
they are therefore rather close to the phenomenological realm
but they are confronted with huge difficulties when
trying to incorporate this physical intuition within a
completely new formalism.
On the other side there are the attempts of
one-step full unification of gravitation and
quantum mechanics,
which usually start from some intuition concerning the
appropriate formalism ({\it e.g.}, canonical/loop
quantum gravity or critical
superstrings)
but are confronted by huge difficulties when
trying to ``come down'' to the level of phenomenological
predictions.
These complementary perspectives might
meet at some mid-way point leading
to new insight in quantum gravity physics.
One instance in which this mid-way-point meeting 
has already been successful is provided by the
mentioned results reported in Ref.~\cite{gampul},
where the
candidate phenomenon of quantum-gravity induced
deformed dispersion relations,
which had been proposed within phenomenological
analyses~\cite{aemn1,gacxt,gacgrb} of the type needed
for the search of a low-energy theory of quantum gravity,
was shown to be consistent with
the structure of canonical/loop quantum gravity.

\subsection{Theories on non-commutative Minkowski
space-time}

At various points in these notes there is a more 
or less explicit reference to deformed symmetries and
noncommutative space-times\footnote{The general idea of some form
of connection between Planck-scale physics and quantum groups 
(with their associated noncommutative geometry)
is of course not new, see {\it e.g.}, 
Refs.~\cite{Ma:pla,Ma:the,Ma:reg,
LNRT:def,kpoin,firenze,maggiore,doplicher,kempfmangano}.
Moreover, some support for noncommutativity of space-time
has also been found within measurability
analyses~\cite{dharam94grf,gacxt}.}.
Just in the previous subsection I have recalled the 
conjecture~\cite{gacxt,gacgrf98}
that an effective large-distance description of some
aspects of quantum gravity might involve quantum symmetries
and noncommutative geometry.
The type of {\it in vacuo} dispersion which can be
tested~\cite{gacgrb} using observations of gamma rays from
distant astrophysical sources is naturally encoded within
a consistent deformation
of Poincar\'e symmetries~\cite{gacxt,lukipapers,gacmaj}.

A useful structure (at least for toy-model purposes,
but perhaps even more than that)
appears to be the noncommutative (so-called ``$\kappa$'')
Minkowski space-time~\cite{firstkappa,shahnkappamin,kpoin}
\begin{eqnarray}
[x^i,t]=\imath\lambda \, x^i,\quad [x^i,x^j]=0
\label{kmink}
\end{eqnarray}
where $i,j=1,2,3$ and $\lambda$ (commonly denoted\footnote{This
author is partly responsible~\cite{gacmaj} for the redundant
convention of using the notation $\lambda$
when the reader is invited to visualize a length scale
and going back to the $\kappa$ notation
when instead it might be natural for the reader to visualize
a mass/energy scale. In spite of its unpleasantness, this redundancy
is here reiterated in order to allow the reader to quickly
identify/interpret corresponding equations in Ref.~\cite{gacmaj}.}
by $1/\kappa$)
is a free length scale.
This simple noncommutative space-time could be taken as a
basis for an effective description of phenomena associated with a
nontrivial foamy quantum-gravity vacuum\footnote{In particular,
within one particular attempt to model space-time foam,
the one of {\it Liouville} non-critical strings~\cite{emn},
the time ``coordinate'' appears~\cite{aemn2} to have properties 
that might be suggestive of a $\kappa$-Minkowski space-time.}.
When probed very softly such a space would appear as
an ordinary Minkowski
space-time\footnote{Generalizations would of course be necessary for
a description of how the quantum-gravity foam affects spaces which are
curved (non-Minkowski) at the classical level, and
even for spaces which are Minkowski at the classical level
a full quantum gravity of course would predict
phenomena which could not be simply encoded in noncommutativity of
Minkowski space.}, but 
probes of sufficiently high energy would be affected by the
properties of the quantum-gravity foam and one could attempt to model
(at least some aspects of) the corresponding dynamics using a
noncommutative Minkowski space-time.
In light of this physical motivation
it is natural to assume that $\lambda$
be related to the Planck length.

The so-called $\kappa$-deformed
Poincar\'e quantum group \cite{LNRT:def}
acts covariantly~\cite{shahnkappamin} on the $\kappa$-Minkowski 
space-time (\ref{kmink}). The dispersion relation
for massless spin-0 particles
\begin{eqnarray}
\lambda^{-2}\left(e^{\lambda E}
+e^{-\lambda E}-2\right)-\vec{k}^2e^{-\lambda E}
=0 ~,
\label{disp}
\end{eqnarray}
which at low energies describes a deformation that is linearly
suppressed by $\lambda$ (and therefore, if indeed $\lambda \sim L_p$, 
is of the type discussed in Section~5),
emerges~\cite{kpoin,lukipapers,gacmaj} as the appropriate Casimir 
of the $\kappa$-deformed Poincar\'e group.
Rigorous support for the 
interpretation of (\ref{disp}) as a {\it bona fide}
dispersion relation characterizing the propagation of waves
in the $\kappa$-Minkowski space-time 
was recently provided in Ref.~\cite{gacmaj}.

In Ref.~\cite{gacmaj} it was also observed that,
using the quantum group
Fourier transform which was worked out for our particular algebra
in Ref.~\cite{MaOec:twi},
there might be a rather simple approach to the definition
of a field theory on the $\kappa$-Minkowski space-time.
In fact, through the quantum group Fourier transform
it is possible to rewrite structures living
on noncommutative space-time as structures living on a classical
(but nonAbelian) ``energy-momentum'' space.
If one is content to evaluate everything in energy-momentum space,
this observation gives the opportunity to by-pass all problems
directly associated with the non-commutativity of space-time.
While waiting for a compelling space-time formulation
of field theories on noncommutative geometries to emerge,
it seems reasonable to restrict all considerations
to the energy-momentum space.
This approach does not work for {\it any} noncommutative
space-time but only for those where the space-time coordinate algebra
is the enveloping algebra of a Lie algebra, with the Lie algebra
generators regarded `up side down' as noncommuting
coordinates~\cite{Ma:ista}.\footnote{Another (partly related,
but different) $\kappa$-Minkowski motivated proposal for
field theory was recently put forward in Ref.~\cite{lukiernew}.
I thank J.~Lukierski for bringing this paper to my attention.}

Within this viewpoint a field theory is not naturally described
in terms of a Lagrangian, but rather it is characterized directly
in terms of Feynman diagrams.
In principle, according to this proposal a given ordinary field theory
can be ``deformed'' into a
counterpart living in a suitable noncommutative
space-time not by fancy quantum-group methods but simply
by the appropriate modification of the momentum-space
Feynman rules to those appropriate for a nonAbelian group.
Additional considerations can be found in Ref.~\cite{gacmaj},
but, in order to give at
least one example of how this nonAbelian deformation
could be applied, let me observe here that the natural propagator
of a massless spin-0 particle on $\kappa$-Minkowski space-time
should be given in energy-momentum space by
the inverse of the operator in the dispersion relation (\ref{disp}), 
i.e. in place of $D=(\omega^2 - {\vec k}^2 - m^2)^{-1}$ one 
would take
\begin{eqnarray}
D_\lambda= \left(\lambda^{-2}
(e^{\lambda \omega} + e^{- \lambda \omega} -2)
- e^{-\lambda \omega} {\vec k}^2 \right)^{-1}
~.
\label{prop}
\end{eqnarray}
As discussed in Ref.~\cite{gacmaj} the elements of this 
approach to field theory appear to lead naturally to a deformation
of CPT symmetries, which would first show up in experiments as a 
violation of ordinary CPT invariance.
The development of realistic field theories of this type
might therefore provide us a single workable
formalism\footnote{Until now
the young field of quantum-gravity phenomenology has relied 
on ``single-use'' phenomenological models (the parameters of the
phenomenological model are only relevant in one physical context).
A first step toward a greater maturity of this phenomenological
programme would be the development of phenomenological models
that apply to more than one physical context (the same parameters 
are fitted using data from more than one physical context).
The type of field theory on $\kappa$-Minkowski space-time
that was considered in Ref.~\cite{gacmaj} (with its single 
parameter $\lambda$) could represent a
first example of these more ambitious multi-purpose
phenomenological models.}
in which  both {\it in vacuo} dispersion and violations of 
ordinary CPT invariance could be computed explicitly (rather than 
being expressed in terms of unknown parameters),
connecting all of the aspects of these candidate
quantum-gravity phenomena to the value
of $\lambda \equiv 1/\kappa$.
One possible ``added bonus'' of
this approach could be associated with
the fact that also loop integration must be 
appropriately deformed, and it appears plausible~\cite{gacmaj}
that (as in other quantum-group based approaches~\cite{Ma:reg})
the deformation might render ultraviolet finite 
some classes of diagrams which would 
ordinarily be affected by ultraviolet divergences.

\section{CONSERVATIVE MOTIVATION AND OTHER \hfil $~$
CLOSING REMARKS}

Since this paper started off with the conclusions,
readers might not be too surprised of the fact
that I devote most of the closing remarks to some
additional motivation.
These remarks had to be postponed until the very end
also because
in reviewing the experiments it would
have been unreasonable to take a conservative viewpoint: those who
are so inclined should find in the
present lecture notes encouragement for
unlimited excitement. However, before closing I must take a step
back and emphasize those reasons of interest in this emerging
phenomenology which can be shared even by those readers who are
approaching all this from a conservative viewpoint.

In reviewing these quantum-gravity experiments I have 
not concealed my (however moderate)
optimism regarding the prospects 
for data-driven advances in quantum-gravity research.
I have reminded the reader of the support one finds in 
the quantum-gravity literature for the type of phenomena which 
we can now start to test, particularly distance fuzziness and
violations of Lorentz and/or CPT symmetries
and
I have also emphasized that it is thanks to recent advances 
in experimental techniques and ideas that these phenomena can be tested
(see, for example,
the role played by the remarkable sensitivities recently
achieved with modern interferometers
in the experimental proposal reviewed in Section~4
and
the role played by very recent break-throughs in GRB phenomenology
in the experimental proposal reviewed in Section~5).
But now
let me emphasize that even
from a conservative viewpoint
these experiments are extremely significant, especially those
that provide tests of quantum mechanics
and tests of fundamental symmetries.
One would not ordinarily need to stress this,
but since these lectures are primarily addressed
to young physics students let me observe that of
course this type of tests is crucial for a sound development
of our science. Even if there was no theoretical argument 
casting doubts on them, we could not possibly take for granted
(extrapolating {\it ad infinitum})
ingredients of our understanding of Nature as crucial 
as its mechanics laws and its symmetry structure.
We should test quantum mechanics and fundamental symmetries
anyway, we might as well do it along the directions which 
appear to be favoured by some quantum-gravity ideas.

One important limitation of the present stage in the
development of quantum-gravity phenomenology
is the fact that most of the experiments
actually test only one of the two main branches of quantum-gravity
proposals: the proposals in which (in one or another fashion)
quantum decoherence is present.
There is in fact a connection (whose careful discussion I postpone
to future publications) between decoherence and the type of 
violations of Lorentz and CPT symmetries and the type of power-law 
dependence on $T_{obs}$ of distance fuzziness here considered.
The portion of our community which finds appealing
the arguments supporting the decoherence-inducing Wheeler-Hawking
space-time foam (and certain views on the 
so-called ``black-hole information paradox'')
can use these recent developments in quantum-gravity phenomenology
as an opportunity for direct tests of some of its intuition.
The rest of our community has developed an
orthogonal intuition concerning the quantum-gravity realm,
in which there is no place for quantum decoherence.
The fact that we are finally at least at the point
of testing decoherence-involving quantum-gravity approaches
(something which was also supposed to be impossible)
should be seen as encouragement for the hope that even 
other quantum-gravity approaches will eventually be 
investigated experimentally.

Even though there is of course no guarantee that this
new phenomenology will be able to
uncover important elements of the structure of quantum gravity,
the fact that such a phenomenological programme exists
suffices to make a legitimate (empirical) science
of quantum gravity, a subject often derided as
a safe heaven for theorists wanting to speculate
freely without any risk of being proven wrong by experiments.
As emphasized in Refs.~\cite{dharamnature,technic}
(and even in the non-technical press~\cite{divulg})
this can be an important turning point in the development of the
field. Concerning the future of quantum-gravity
phenomenology let me summarize my expectations
in the form of a response to the question posed by
the title of these notes:
I believe that we are indeed at the dawn
of quantum-gravity phenomenology, but the forecasts
call for an extremely long and cloudy day with only a few rare
moments of sunshine.
Especially for those of us motivated by theoretical arguments
suggesting that at the end of the road there should be 
a wonderful revolution of our understanding of Nature
(perhaps a revolution of even greater magnitude than the one 
undergone during the first years of this 20th century),
it is crucial to
profit fully from
the few glimpses of the road ahead which
quantum-gravity phenomenology will provide.

\vglue 0.6cm
\leftline{\Large {\bf Acknowledgements}}
\vglue 0.4cm
First of all I would like to thank the organizers
of this XXXV Winter School, especially for their role
in creating a very comfortable informal atmosphere, which
facilitated exchanges of ideas among the participants.
My understanding of Refs.~\cite{elmn}, \cite{stringcogwi}
and \cite{grwlarge}
benefited from conversations 
with R.~Brustein, M.~Gasperini,
G.F.~Giudice, N.E.~Mavromatos, G.~Veneziano and J.D.~Wells, 
which I very greatfully acknowlegde.
Still on the ``theory side'' I am grateful to several colleagues 
who provided encouragement and stimulating feed-back, 
particularly D.~Ahluwalia, A.~Ashtekar,
J.~Ellis, J.~Lukierski,
C.~Rovelli, S.~Sarkar, L.~Smolin and J.~Stachel.
On the ``experiment side'' I would like to thank
F.~Barone, 
E.~Bloom,
J.~Faist, 
R.~Flaminio,
L.~Gammaitoni, 
T.~Huffman,
L.~Marrucci,
M.~Punturo
and
J.~Scargle
for informative conversations (some were nearly tutorials)
on various aspects of interferometry
and gamma-ray-burst experiments.

\bigskip
%\newpage

\baselineskip 12pt plus .5pt minus .5pt

\end{document}